\documentclass[reprint,superscriptaddress,amsmath,amssymb,aps,preprintnumbers]{revtex4-2}
\usepackage{graphicx}
\usepackage{caption}
\usepackage{dcolumn}
\usepackage{bm}
\usepackage[usenames]{color}
\usepackage[normalem]{ulem}
\usepackage{verbatim}

\RequirePackage{graphicx,epstopdf}
\usepackage[colorlinks=true, pdfstartview=FitV, linkcolor=blue, citecolor=blue, urlcolor=blue]{hyperref}
\graphicspath{{./Figs/}}

\renewcommand{\sout}[1]{\bgroup \color{red} \ULdepth=-.5ex \ULset {#1}}
\newcommand{\refeq}[1]{(\ref{#1})}
\newcommand{\average}[1]{\ensuremath{\langle#1\rangle}}
\newcommand{\bx}{ \bm{x} }
\newcommand{\nf}{ N_{\rm f} }
\newcommand{\nft}{ N^{\rm tot}_{\rm f} }

\begin{document}
	\preprint{}
	\title{
		Quantum simulation of QC$_2$D on a 2-dimensional small lattice
	}
	\author{Zhen-Xuan Yang}
	\email{24110190097@m.fudan.edu.cn}
	\affiliation{Physics Department and Center for Particle Physics and Field Theory, Fudan University, Shanghai 200438, China}
	\author{Hidefumi Matsuda}
	\email{da.matsu.00.bbb.kobe@gmail.com}
	\affiliation{Zhejiang Institute of Modern Physics, Department of Physics, Zhejiang University, Hangzhou, 310027, China}
	\author{Xu-Guang Huang}
	\email{huangxuguang@fudan.edu.cn}
	\affiliation{Physics Department and Center for Particle Physics and Field Theory, Fudan University, Shanghai 200438, China}
	\affiliation{Key Laboratory of Nuclear Physics and Ion-beam Application (MOE), Fudan University, Shanghai 200433, China}
	\affiliation{Shanghai Research Center for Theoretical Nuclear Physics, National Natural Science Foundation of China and Fudan University, Shanghai 200438, China}
	\author{Kouji Kashiwa}
	\email{kashiwa@fit.ac.jp}
	\affiliation{Department of Computer Science and Engineering, Faculty of Information Engineering, Fukuoka Institute of Technology, Fukuoka 811-0295, Japan}
	
	\begin{abstract}
		We study the Hamiltonian formulation of SU(2) Yang-Mills theory with staggered fermions in a (2+1)-dimensional small lattice system. We construct a gauge-invariant and finite-dimensional Hilbert space for the theory by applying the loop-string-hadron formulation and specifically map the model to a spin system. We classically emulate digital quantum simulation and observe the real-time evolution of the single-site entanglement entropy, the fermion entanglement entropy, and the fermion pair production.
	\end{abstract}
	\maketitle
	\section{Introduction}
	Gauge theory plays a crucial role in constructing the framework of modern particle physics. 
	In particular, the SU(3) gauge theory provides the foundation for Quantum Chromodynamics (QCD) that describes the strong interaction among quarks and gluons. 
	One important feature of QCD is asymptotic freedom, which implies that perturbation theory becomes invalid at low energy scales.
	Therefore, researchers have developed several non-perturbative approaches to investigate the low-energy physics of QCD. 
	One of the most successful approaches is the first-principles calculation based on the path integral formulation for discrete variables, known as lattice QCD~\cite{Wilson:1974sk}.
	
	By numerically evaluating the path integral and taking the continuum limit, one can, in principle, investigate thermal expectation values of observables even in the non-perturbative region.
	
	Conventionally, this integration is performed using the Markov-Chain Monte Carlo method, which does not work when the
	integrand is non-real positive and highly oscillating.
	This kind of difficulty in numerical integration is known as the sign problem~\cite{PhysRevLett.94.170201,deForcrand:2009zkb,Gattringer:2016kco,Nagata:2021bru,*Nagata:2021ugx}, which occurs in a wide range of fields in physics. 
	In lattice QCD calculations, one encounters the sign problem in situations such as those involving topological effects induced by the theta term, the finite baryon chemical potential region, and real-time evolution.
	
	Even though there have been various approaches to challenge the sign problem based on the Lagrangian formulism, such as the complex Langevin method~\cite{Parisi:1980ys,Klauder:1983sp,Parisi:1983mgm}, Lefschetz thimble method~\cite{Witten:2010cx,Witten:2010cx,Cristoforetti:2012su,Fujii:2013sra} and some related methods~\cite{Mori:2017pne,Mori:2017nwj,Alexandru:2018fqp,Namekawa:2022liz}, their success has yet to lead to practical applications in QCD.
	
	The Hamiltonian formulation on a lattice has recently gained attention as an alternative approach to studying gauge theories~\cite{Kogut:1974ag,Kogut:1979wt}. This method avoids the sign problem by directly dealing with quantum states in their Hilbert space and calculating observables, rather than relying on Monte Carlo integration. However, the Hamiltonian formulation introduces a different practical challenge: the vastness of the Hilbert space to be handled. The Hilbert space of gauge fields is infinitely large because they are bosonic theories. Therefore, in practice, it is necessary to truncate the theories to make their Hilbert space finite.
	Several truncation methods have been proposed, including the loop-string-hadron (LSH) approach~\cite{Raychowdhury:2019iki,Raychowdhury:2018osk,Davoudi:2020yln}, q-deformation~\cite{Zache:2023dko,Hayata:2023bgh,Hayata:2023puo}, local multiplet basis~\cite{Klco:2019evd,Ciavarella:2021nmj}, and quantum link model~\cite{Banerjee:2017tjn}.
	Each formulation has its own distinct features. None of the approaches are currently applicable to 3+1 dimensional QCD. Researchers typically start with simpler cases than QCD, such as lower spatial dimensions, fewer flavors and colors, and by neglecting fermionic degrees of freedom.
	
	Nevertheless, even after truncation, the Hilbert space often remains significantly large. To address this issue, quantum computation emerges as a promising tool. A quantum computer with $N$ qubits has a memory capacity of $O$($2^N$), whereas a classical computer with $N$ classical bits has a memory capacity of only $O$($N$). 
	Due to the development of quantum computing, it is convincing that quantum computer will be able to deal with this huge Hilbert space in the future as its computing resource increases exponentially with the increase in the number of qubits.

	With the development in the formulation, significant progress has also been made in numerical simulations of gauge theories. 
	Although numerous efforts have been made on investigating the quantum simulation of lattice gauge theories, 
	there are still many steps to bridge the gap to QCD.
	For example, many studies focus on 1+1 dimensional physics, such as the Schwinger model~\cite{Byrnes:2002nv,Banuls:2013jaa,Martinez:2016yna,Muschik:2016tws,Klco:2018kyo,Kokail:2018eiw,Farrell:2023fgd,Farrell:2024fit,Chen:2024pee}. 
	In the higher spatial dimensions such as 2+1D and 3+1D, there are studies investigating the dynamics of non-Abelian gauge theory without fermions~\cite{Klco:2019evd,Ciavarella:2021nmj,ARahman:2021ktn,ARahman:2022tkr,Hayata:2020xxm,Hayata:2021kcp,Zache:2023dko,Hayata:2023bgh,Hayata:2023puo,Yao:2023pht,Muller:2023nnk,Ebner:2023ixq,Turro:2024pxu,Ebner:2024mee,Turro:2025sec,Illa:2025dou}.
    
	However, simulations of non-Abelian gauge theory with fermions in more than two spatial dimensions are still lacking, making it an open and attractive question. 
	
	One of the most interesting applications of the model is its real-time quantum evolution to investigate the thermalization of SU(2) gauge theory with fermions. This problem is closely related to the understanding of the thermalization process in the early stages of relativistic heavy-ion collisions~\cite{Heinz:2001xi,Berges:2020fwq}.
	The theoretical picture of relativistic heavy-ion collisions based on perturbative QCD suggests that the initial state of the bulk matter produced is a far-from-equilibrium system of gluons~\cite{McLerran:1993ni,McLerran:1993ka,McLerran:1994vd,Kovner:1995ja}. This gluonic matter thermalizes due to the intrinsic dynamics of QCD, while fermions are produced, eventually evolving into a quark-gluon plasma (QGP). However, our understanding of the nonequilibrium processes remains incomplete due to the numerical difficulty of simulating the real-time nonequilibrium evolution of QCD.
	
	In recent years, significant progress has been made in understanding quantum thermalization, particularly in the field of condensed matter physics. It has been suggested that quantum thermalization can be explained by the eigenstate thermalization hypothesis (ETH), which is also known to be closely related to entanglement and quantum chaos~\cite{PhysRevA.43.2046,PhysRevE.50.888,Rigol:2007juv}. Motivated by these developments, increasing attention has been given to investigating thermalization mechanisms in quantum gauge theories using Hamiltonian-based simulations, even in finite, low-dimensional SU(2) systems~\cite{Hayata:2020xxm,Yao:2023pht,Ebner:2023ixq,Ebner:2024mee,cataldi2024simulating}.
	Building upon this research direction, investigating the thermalization of SU(2) gauge theory by incorporating fermions is a worthwhile endeavor. By including fermions, we can further explore fermion production and fermion entropy.
	
	In this paper, we study the real-time quantum evolution of the SU(2) Yang-Mills theory with fermions in a 2+1 dimensional small isolated system by using the LSH formulation.
	We consider the Kogut-Susskind Hamiltonian formulation~\cite{Kogut:1974ag,Kogut:1979wt} on a single square lattice and introduce staggered fermions~\cite{Susskind:1976jm,Kogut:1979wt} on the vertexes to circumvent the fermion doubling~\cite{Nielsen:1980rz,Nielsen:1981xu}. 
	Using the Schwinger boson representation~\cite{Mathur:2004kr,Raychowdhury:2019iki} of the SU(2) gauge field, we can
	construct the physical states on the lattice by solving the Gauss law and figure out the actions of the Hamiltonian~\cite{Mathur:2004kr,Raychowdhury:2019iki}.
	Our simulation is based on a quantum computer emulator~\cite{Raychowdhury:2019iki,P:2019qdq}. To perform the quantum simulation, we map the model to a spin system. 
	
	This paper is organized as follows. In Sec.~\ref{Sec2}, we first review the Kogut-Susskind Hamiltonian and show the necessary formulations. Then, we introduce the construction of the physical states by applying gauge invariant constraints and specifically calculate actions of the Hamiltonian. In Sec.~\ref{Sec3}, we show how to map the model to a spin system and discuss our numerical results which mainly focus on entanglement and pair production. Sec.~\ref{Sec4} is summary and outlook. In Appendix \ref{A2}, we show the actions of all necessary gauge invariant operators. In Appendix \ref{B}, we show the exact identity between the lattice system and the spin system. Besides, we also show all terms of the Hamiltonian matrix.

	\section{HAMILTONIAN FORMULATION}\label{Sec2}

	\subsection{Formulation}
	
	We review the Hamiltonian formulation of the SU(2) lattice Yang-Mills theory, 
	which is so-called Kogut-Susskind Hamiltonian ~\cite{Kogut:1974ag,Kogut:1979wt}, 
	on a two-dimensional spatial lattice with the staggered lattice fermion.
	The Hamiltonian is given as  
	\begin{align}\label{Hemf}
		&H=H_{\rm E} + H_{\rm B} + H_{\rm F}\ ,
	\end{align}
	where
	\begin{align}
		H_{\rm E} &=  K\sum_{\boldsymbol{x}, \mu}  E^{2}(\boldsymbol{x}, \mu) ,\\
		H_{\rm B} &= -\frac{1}{2K} \sum_{P} \operatorname{tr}\left[U_{\mu}(\boldsymbol{x}) U_{\nu}\left(\boldsymbol{x}+\hat{e}_{\mu}\right) U_{\mu}^{\dagger}\left(\boldsymbol{x}+\hat{e}_{\nu}\right) U_{\nu}^{\dagger}(\boldsymbol{x})\right] \nonumber\\
		&+ \text { (h.c.) } ,\\
		H_{\text {F}}&= \sum_{\mu, x}
		\Bigl[\eta_{\mu}(x)\left\{\psi^{\dagger}(\boldsymbol{x}) U_{\mu}(\boldsymbol{x}) \psi\left(\boldsymbol{x} +\hat{e}_{\mu}\right)+(\text { h.c. })\right\}
		\nonumber\\
		&-(-1)^{x_{1}+x_{2}} m \psi^{\dagger}(\boldsymbol{x}) \psi(\boldsymbol{x})\Bigr]\ ,\label{Eq:Hk}
	\end{align}
	here $\eta_{1}(\boldsymbol{x})=(-1)^{x_{1}+x_{2}}$ and $\eta_{2}(\boldsymbol{x})=(-1)^{x_{2}}$. 
	In this Hamiltonian,  $E(\boldsymbol{x}, \mu)$, defined on the ends of links, is the chromo-electric field and $U_{\mu}(\boldsymbol{x})$ 
	is the link variable which is defined on the link. 
	Besides, $\psi(x)$ is the staggered fermion field which is defined on the sites. 
	Under this definition, $H_{\rm E},H_{\rm B}$ and $H_F$ are the electric part, 
	magnetic part and fermionic part of the Hamiltonian, respectively. 
	In this formula, $\boldsymbol{x}$ and $\mu(=x^1, x^2)$ respectively represent 
	the site position of $2$-dimensional spatial lattice and directions with 
	$\hat{e}_{\mu}$ being the unit vector along the direction $\mu$. 
	Besides, $P$ denotes the plaquette. 
	Note that we redefine the Hamiltonian, shown in Eq.~\refeq{Hemf}, by multiplying a factor $2a$ with $a$ being the lattice constant. 
	According to this, the parameter $K$ equals to $g^2$ with $g$ being the coupling constant 
	and the mass parameter, $m$, equals to $2m_0 a$ with $m_0$ being the intrinsic mass of fermion field.
	However, throughout this paper, unless otherwise specified, other physical quantities are normalized by the lattice spacing $a$.
	
	There are two types of representations of the chromo-electric field $E(\boldsymbol{x}, \mu)$ on the lattice.
	One is the left electric field $E_{\rm L}^{a}(\boldsymbol{x}, \mu)$, located at position $\bx$ on the edge of the link connecting $x$ to $x+\hat{\mu}$. The other is the right electric field $E_{\rm R}^{a}(\boldsymbol{x}, \mu)$, located at position $\bx$ on the edge of the link connecting $x-\hat{\mu}$ to $x$.
	For the lattice SU(2) Yang-Mills theory, the non-vanishing commutation relation among the link variable and electric fields are given as  
	\begin{align}
		{\left[E_{\rm L}^{a}(\boldsymbol{x}, \mu), U_{\nu}(\boldsymbol{y})\right] } & =-\frac{\sigma^{a}}{2}  U_{\mu}(\boldsymbol{x}) \delta_{\mu \nu} \delta_{\boldsymbol{x} \boldsymbol{y}} \label{Cre1}\ ,\\
		{\left[E_{\rm R}^{a}(\boldsymbol{x}, \mu), U_{\nu}(\boldsymbol{y})\right] } & =U_{\mu}(\boldsymbol{x}) \frac{\sigma^{a}}{2} \delta_{\mu \nu} \delta_{\boldsymbol{x} \boldsymbol{y}} \label{Cre2}\ ,\\
		{\left[E_{\rm L}^{a}(\boldsymbol{x}, \mu), E_{\rm L}^{b}(\boldsymbol{y}, \nu)\right] } & =\mathrm{i} \epsilon^{a b c} E_{\rm L}^{c}(\boldsymbol{x}, \mu) \delta_{\mu \nu} \delta_{\boldsymbol{x} \boldsymbol{y}}\label{Cre3}\ ,\\
		{\left[E_{\rm R}^{a}(\boldsymbol{x}, \mu), E_{\rm R}^{b}(\boldsymbol{y}, \nu)\right] } & =\mathrm{i} \epsilon^{a b c} E_{\rm R}^{c}(\boldsymbol{x}, \mu) \delta_{\mu \nu} \delta_{\boldsymbol{x} \boldsymbol{y}}\label{Cre4}\ ,
	\end{align}
	where $\sigma^{a=1,2,3}$ are the Pauli matrices, $\delta_{\mu \nu}$ is the Kronecker delta, 
	and $\epsilon^{a b c}$ is the Levi-Civita symbol with $\epsilon^{123}=1$. 
	These relations mean that the left and right electric fields serve as generators for the link variable in the SU(2) gauge transformation.
	Note that we have the equality condition between the two types of electric fields as
	\begin{equation}\label{U1}
		\sum_{a} E_{\rm L}^{a}(\boldsymbol{x}, \mu) E_{\rm L}^{a}(\boldsymbol{x}, \mu) 
		=\sum_{a} E_{\rm R}^{a}(\boldsymbol{x}, \mu) E_{\rm R}^{a}(\boldsymbol{x}, \mu)\ ,
	\end{equation}
	indicating that the electric part of the Hamiltonian, given in Eq.~\refeq{Hemf}, does not depend on the representation used.
	The physical states $|\Psi\rangle$ defined on the sites should satisfy the Gauss law constraint: 
	\begin{equation}\label{Glaw}
		\left(\sum_{\mu}\left[E_{\rm L}^{a}(\boldsymbol{x}, \mu)+E_{\rm R}^{a}\left(\boldsymbol{x}-\hat{e}_{\mu}, \mu\right)\right]+Q^a(\boldsymbol{x})\right)|\Psi\rangle=0\ ,
	\end{equation}
	where the charge operator $Q^a$ is defined as $Q^a(\boldsymbol{x})=\psi^{\dagger}(\boldsymbol{x}) \frac{\sigma^{a}}{2} \psi(\boldsymbol{x})$. 
	This charge operator also satisfies the SU(2) algebra as
	\begin{equation}\label{chargerel}
		{\left[Q^{a}(\boldsymbol{x}), Q^{b}(\boldsymbol{y})\right] } =\mathrm{i} \epsilon^{a b c} Q^{c}(\boldsymbol{x}) \delta_{\boldsymbol{x} \boldsymbol{y}}\ .
	\end{equation}
	
	To construct a gauge-invariant basis for the Hilbert space and to calculate the matrix elements of observables in this basis, 
	we use the Schwinger boson representation of the gauge field operator~\cite{Mathur:2004kr,Raychowdhury:2019iki}.
	The electric field can be rewritten as
	\begin{align}
		E^a_{\rm L}=\alpha^{\dagger}_i\frac{\sigma^a_{ij}}{2}\alpha_j\ ,\\
		E^a_{\rm R}=\beta ^{\dagger}_i\frac{\sigma^a_{ij}}{2}\beta_j\ ,
	\end{align}
	where $\alpha_{i=1,2}$ and $\beta_{i=1,2}$ are both boson annihilation operators whose specific meaning will be given in next subsection. 
	For magnetic field, we define the link variable as $ U=U_{\rm L}(\alpha) U_{\rm R}(\beta)$ with 
	\begin{align}
		U_{\rm L}(\alpha) & =\frac{1}{\sqrt{N_{\rm L}+1}}\left(\begin{array}{cc}
			\alpha_{2}^{\dagger} & \alpha_{1} \\
			-\alpha_{1}^{\dagger} & \alpha_{2}
		\end{array}\right)\ ,\\
		U_{\rm R}(\beta) & =\left(\begin{array}{cc}
			\beta_{1}^{\dagger} & \beta_{2}^{\dagger} \\
			-\beta_{2}           & \beta_{1}
		\end{array}\right) \frac{1}{\sqrt{N_{\rm R}+1}}\ ,
	\end{align}
	which is the well-known form, reproducing the commutation relation of gauge field operators.
	
	As the Fock space for Schwinger bosons, we can define the Hilbert space.
	Note that the defined Hilbert space is larger than the Hilbert space for the original theory even after limiting it to the subspace that satisfies the Gauss law.
	The equality condition (\ref{U1}) indeed imposes an additional constraint on the physical state, in addition to the Gauss law, as
	\begin{equation}\label{eqc}
		N_{\rm L}(\boldsymbol{x}, \mu)|\Psi_{\rm L}\rangle=N_{\rm R}(\boldsymbol{x}, \mu)|\Psi_{\rm R}\rangle\ ,
	\end{equation}
	where $N_{\rm L}=\sum_{i} \alpha^\dagger_i \alpha_i$ and $N_{\rm R}=\sum_{i} \beta^\dagger_i \beta_i$ 
	are the number operators of Schwinger bosons. 
	Here $|\Psi_{\rm L}\rangle,|\Psi_{\rm R}\rangle$ are the physical states located at the left and right sites of the link $(\boldsymbol{x}, \mu)$. 
	This implies that the number of Schwinger bosons on both sides of one link must be the same. 
	This constraint is called the Abelian Gauss law~\cite{Raychowdhury:2019iki}.

				Based on the above formulas, we can calculate the real-time evolution of the 2-dimensional square lattice system according to the Schr\"odinger equation. 
				But before this, we need to construct the physical state $|\Psi\rangle$ to satisfy
				the Gauss law, shown in Eq.~\refeq{Glaw}, and the equality condition, shown in Eq.~\refeq{eqc}.

				\subsection{Construction of physical state}\label{22}
				
				\begin{figure*}[tp]
					\begin{minipage}{0.3\textwidth}
						\includegraphics[width=1\textwidth]{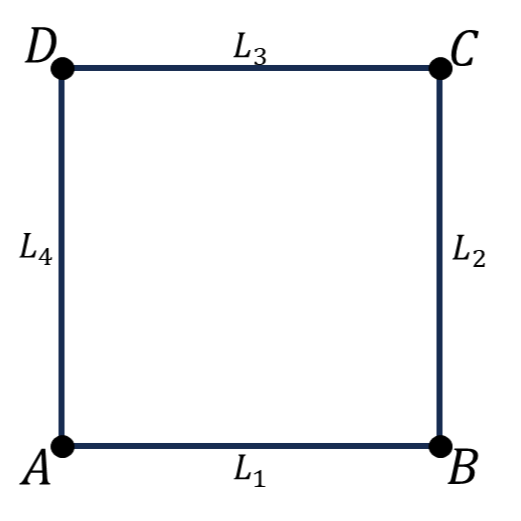}
					\end{minipage}
					\hspace{7.5em}
					\begin{minipage}{0.3\textwidth}
						\includegraphics[width=\textwidth]{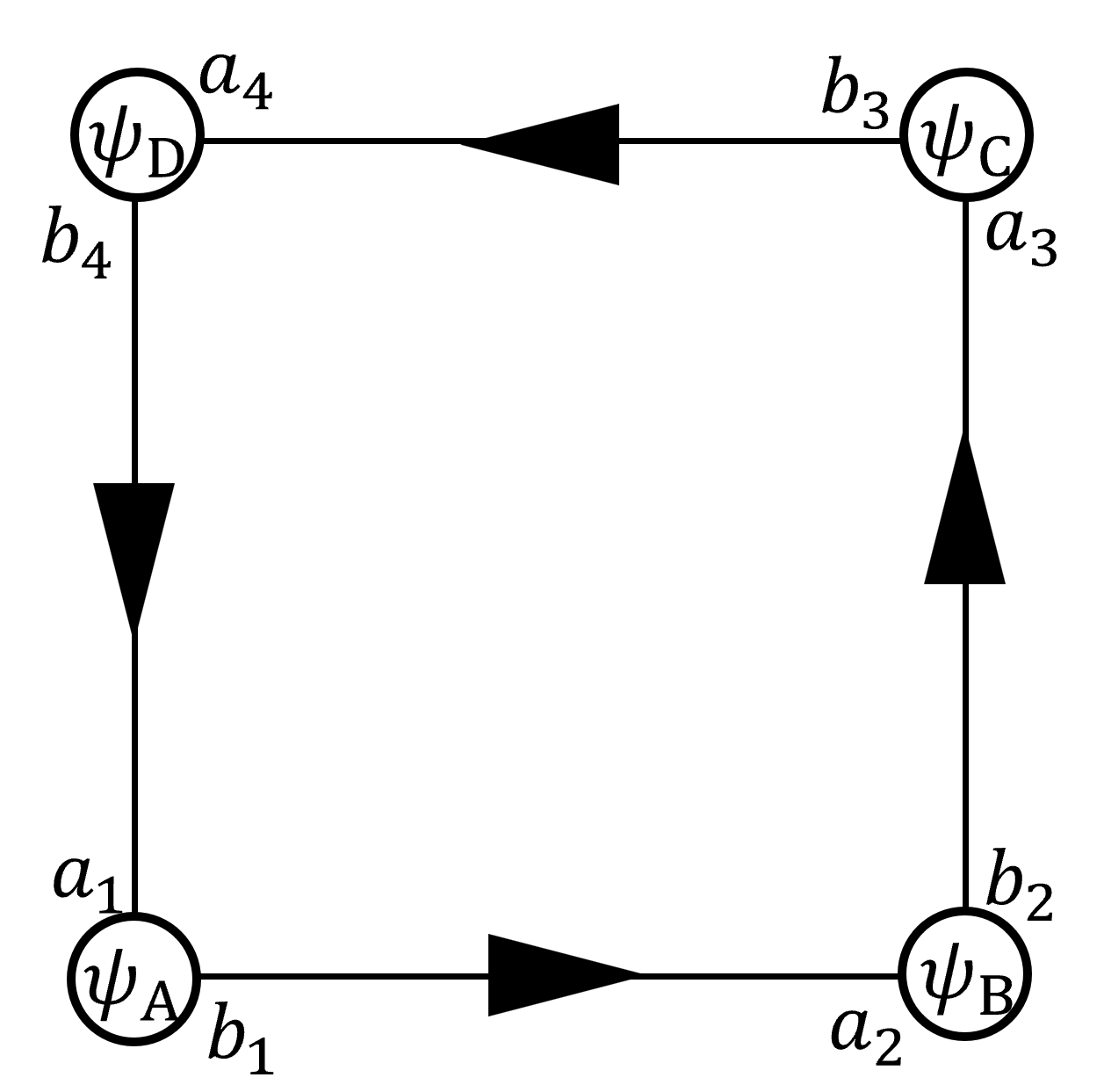}
					\end{minipage}
					\caption{
						Left:  
						Labels of links and sites on the (2+1) dimension small lattice model.
						The links are labeled as $L_i$ with $i=1,2,3$, and $4$.
						The sites are labeled as A, B, C, and D, each representing the site at 
						$(x^1, x^2) = (0, 0), (1, 0), (1, 1),$ and $(0, 1)$, respectively.
						Right:
						Labels of the boson and fermion operators on sites on the (2+1) dimension small lattice model.
						Along a path moving counterclockwise on the square lattice, we define the bosons associated with a given lattice point in the order, 
						$b \to a$.
					}
					\label{Fig:plaquette}
				\end{figure*}
				In this section, we intend to construct the physical state on the (2+1) dimensional little lattice. 
				Basically, we have two constraints: the Gauss law constraint (\ref{Glaw}) for a local state at a single site and the equality condition (\ref{eqc}) for connecting two adjacent sites. 
				The labels for links and sites are given in the left of Fig.~\ref{Fig:plaquette}.
				The labels for the annihilation operators of bosons and fermions at the sites are shown on the right in Fig.~\ref{Fig:plaquette}. 
				The labels for the creation operators are obtained by adding a dagger symbol to each annihilation operator, such as $a^\dagger_1, b^\dagger_1$ and $\psi^\dagger_A$.
				At this point, we do not distinguish whether the boson operator represents the left or right electric field, 
				or whether the fermion operator represents a fermion or anti-fermion. 
				However, later, we see that the fermions at sites B and D represent fermions, while those at sites A and C represent anti-fermions.
				
				First, let us focus on the local states at single sites that preserve Gauss law.
				Such a state can be expressed by gauge-invariant operators as follows~\cite{Raychowdhury:2019iki}:
				\begin{align}
					&|s,r_1,r_2,v\rangle \nonumber\\
					=&
					\frac{1}{\sqrt{C_{\text {n }}}}\left(\boldsymbol{b}^{\dagger} \cdot \tilde{\boldsymbol{a}}^{\dagger}\right)^{s}\left(\boldsymbol{b}^{\dagger} \cdot \tilde{\boldsymbol{\psi}}^{\dagger}\right)^{r_{1}}\left(\boldsymbol{a}^{\dagger} \cdot \tilde{\boldsymbol{\psi}}^{\dagger}\right)^{r_{2}}\left(\boldsymbol{\psi}^{\dagger} \cdot \tilde{\boldsymbol{\psi}}^{\dagger}\right)^{v}
					|0\rangle\ ,
					\label{Eq:localstate}
				\end{align}
				where 
				$s,r_1,r_2,v$ denote the quantum number of the state, 
				$(a^\dagger,b^\dagger)$ and $\psi^\dagger$ denote the creation operators of Schwinger bosons and Staggered fermion respectively, 
				$|0\rangle$ is the ground state of the Fock space defined by these operators, 
				and $C_{\text {n }}=4^{v} s!(s+r_{1}+r_{2}+1)$ is the normalization factor. 
				Besides, the dot means inner product in color space with $\tilde{\boldsymbol{a}}=i \sigma^2 \boldsymbol{a} $.
				The parameter $s$ can be any natural number, while $r_1,r_2,v$ can only be 0 or 1 because of the anti-commutation relation of fermion operator.
				We have two useful relation as 
				\begin{align}\label{usefuleq}
					&\left(\boldsymbol{b}^{\dagger} \cdot \tilde{\boldsymbol{\psi}}^{\dagger}\right)\left(\boldsymbol{a}^{\dagger} \cdot \tilde{\boldsymbol{\psi}}^{\dagger}\right)=\frac{1}{2}\left(\boldsymbol{b}^{\dagger} \cdot \tilde{\boldsymbol{a}}^{\dagger}\right)\left(\boldsymbol{\psi}^{\dagger} \cdot \tilde{\boldsymbol{\psi}}^{\dagger}\right)\ ,\\
					&\left(\boldsymbol{a}^{\dagger} \cdot \tilde{\boldsymbol{\psi}}^{\dagger}\right)\left(\boldsymbol{\psi}^{\dagger} \cdot \tilde{\boldsymbol{\psi}}^{\dagger}\right)=\left(\boldsymbol{b}^{\dagger} \cdot \tilde{\boldsymbol{\psi}}^{\dagger}\right)\left(\boldsymbol{\psi}^{\dagger} \cdot \tilde{\boldsymbol{\psi}}^{\dagger}\right)=0\ ,
				\end{align}
				which indicate that states with $r_1=r_2=1, v=0$ are equivalent to states with $r_1=r_2=0, v=1$ and thus any two parameters of $r_1,r_2,v$ can not be 1 at the same time. The expression shown in Eq.~\refeq{Eq:localstate} 
				is symmetric with respect to the exchange of $a$ and $b$.
				The local state, given in Eq.~\refeq{Eq:localstate}, satisfies the following Gauss condition,
				\begin{equation}
					\left( E_a+E_b+Q_\psi \right)|\Psi\rangle=0\ ,\label{Eq:GL}
				\end{equation}
				with 
				$E_a = \bm{a}^{\dagger} \frac{\sigma}{2} \bm{a}$, 
				$E_b = \bm{b}^{\dagger} \frac{\sigma}{2} \bm{b}$, 
				and 
				$Q_\psi = \bm{\psi}^{\dagger} \frac{\sigma}{2} \bm{\psi}$.
				This method to construct the local gauge-invariant state is known as the LSH formulation~\cite{Raychowdhury:2019iki}. 

				Let us consider the local state given in Eq.~\refeq{Eq:localstate} in terms of another representation to discuss.
				Using the fact that the electric field and charge operators all satisfy the SU(2) algebra, the Gauss law shown in Eq.~\refeq{Eq:GL} can be interpreted as the singlet condition of the physical state, namely requiring the ``total angular momentum," $E_a+E_b+Q_\psi$, to vanish.
				Such a singlet state can be expressed in terms of angular momentum quantum numbers as $|j_{\mathrm{b}}, j_{\mathrm{a}}, j_{\mathrm{\psi}}\rangle$, where $j_{\mathrm{b}}, j_{\mathrm{a}}$, and $j_{\mathrm{\psi}}$ denote the angular momentum quantum numbers associated with $E_a, E_b$, and $Q_\psi$, respectively.
				However, the theory has a conserved quantity: the net fermion number. 
				As a result, the fermion number is required as an additional label to distinguish states with different fermion numbers and uniquely specify the physical state.
				This is reflected in the relationship between the angular momentum and quantum numbers for the states in the LSH formulation,
				\begin{equation}
					j_\mathrm{b}=\frac{1}{2}\left(s+r_{1}\right), \quad j_\mathrm{a}=\frac{1}{2}\left(s+r_{2}\right), \quad j_{\mathrm{\psi}}=\frac{1}{2}\left(r_{1}+r_{2}\right)\ ,   
				\end{equation}
				which shows that different values of the label "$v$" can result in the same angular momentum.

				Since the Hilbert space of a bosonic theory is infinite, 
				we need to introduce some truncation to do the quantum simulation practically.
				Here, we only treat local states where all the quantum numbers of the angular momentum are lower than $j_{\text{max}}=\frac{1}{2}$.
				This truncation acts as a cutoff that removes high-energy states.
				This is particularly evident in the electric field part of the Hamiltonian, which can be expressed as a sum of total
				angular momenta. 
				Therefore, we have 6 local states based on Eq.~\refeq{Eq:localstate} as
				\begin{align}
					&|0,0,0,0\rangle\ , |1,0,0,0\rangle\ , \\
					&|0,0,1,0\rangle\ , |0,1,0,0\rangle\ , \\
					&|0,0,0,1\rangle\ , |1,0,0,1\rangle\ .
				\end{align}
				Here states at the first, second and third lines have different eigen values for the number operator, $N_{\rm F} \equiv \psi^\dagger \psi$, 
				each having $\nf=0, 1$ and $2$, respectively.
				To simplify the notation, we relabel these states as  
				\begin{equation}
					\begin{aligned}
						&|0\rangle       = |0,0,0,0\rangle\ ,\\
						&|2\rangle       = |0,0,0,1\rangle\ ,\\
						&|1\rangle       = |0,0,1,0\rangle\ ,\\
						&|\hat{1}\rangle = |0,1,0,0\rangle\ ,\\
						&|\hat{0}\rangle = |1,0,0,0\rangle\ ,\\
						&|\hat{2}\rangle = |1,0,0,1\rangle\ ,
					\end{aligned}
				\end{equation}
				where the left hand side is our new notation with the number being $\nf$, 
				and the hat means that the local state has a finite number of $b$ boson, $\average{b^\dagger b} \neq 0$.
				It should be noted that the number operator $\nf$ is not the fermion number operator, 
				as there are no fermions but rather anti-fermions at sites A and C. 
				Therefore, at sites A and C, $\nf-2$ serves as the fermion number operator. 
				This implies that the total fermion number operator across all sites, $\nft \equiv N_{{\rm f},A}+N_{{\rm f},B}+N_{{\rm f},C}+N_{{\rm f},D}$,
				only differs from the net fermion number by the constant 4, and thus it is conserved. 
				
				For global state, the constraint (\ref{eqc}) implies
				\begin{equation}\label{connect}
					(s+r_{1})_{i}=(s+r_{2})_{f}\ ,
				\end{equation}
				where the $i,f$ mean the start and end of the link along the direction of counterclockwise around the square lattice.
				Thus, we can get all possible two point neighbor states as     
				\begin{equation}\label{neighbor}
					\begin{aligned}
						&|00\rangle ,|02\rangle ,|20\rangle ,|22\rangle,
						|\hat{0}\hat{0}\rangle,|\hat{0}\hat{2}\rangle,|\hat{2}\hat{0}\rangle,|\hat{2}\hat{2}\rangle ,
						|0\hat{1}\rangle\ , \\&|2\hat{1}\rangle,|\hat{0}1\rangle,|\hat{2}1\rangle,
						|10\rangle,|12\rangle,|\hat{1}\hat{0}\rangle,|\hat{1}\hat{2}\rangle,
						|1\hat{1}\rangle ,|\hat{1}1\rangle \ ,
					\end{aligned}
				\end{equation}
				the left and right labels in the ket states correspond to the states at the neighboring sites, respectively.
				Based on the two point states above, we can spell out all global states like $|2020\rangle$,
				where the labels in the state are ordered as (A, B, C, D) from left to right.
				
				To show the global states, we divide them into different groups according to the different eigenvalue of the total number operator 
				$\nft$.
				This is a benefit for both calculation and understanding the physics because of the fermion number conservation. 
				First, we have the states with $\nft = 0$ as
				\begin{align}
					|0000\rangle,\ |\hat{0}\hat{0}\hat{0}\hat{0}\rangle\ .
				\end{align}
				There is no state which has odd fermion number. 
				Thus, next we show the $\nft=2$ states as
				\begin{equation}
					\begin{aligned}
						|2000\rangle,\ |\hat{2}\hat{0}\hat{0}\hat{0}\rangle,
						\ |\hat{1}100\rangle,\ |1\hat{1}\hat{0}\hat{0}\rangle,
						\ |\hat{1}\hat{0}10\rangle\ .
					\end{aligned}
				\end{equation}
				Here we only list representative states because cyclic permutations of labels in any global state automatically give us other global states.
				Thus, the number of  states with $\nft=2$ should be $20$. 
				We list states with $\nft=4$ in the same manner as
				\begin{equation}
					\begin{aligned}
						|\hat{1}1\hat{1}1\rangle,\ |1\hat{1}1\hat{1}\rangle,
						\ |2200\rangle,\ |2020\rangle,
						\ |\hat{2}\hat{2}\hat{0}\hat{0}\rangle,\ |\hat{2}\hat{0}\hat{2}\hat{0}\rangle\ ,\\
						\ |\hat{1}120\rangle,\ |1\hat{1}\hat{2}\hat{0}\rangle,  
						\ |\hat{1}\hat{2}10\rangle,\ |12\hat{1}\hat{0}\rangle,
						\ |\hat{1}102\rangle,\ |1\hat{1}\hat{0}\hat{2}\rangle\ ,\label{Eq:state4}
					\end{aligned}
				\end{equation}
				which gives that the number of states with $\nft=4$ is 38 after cyclic permutation of labels. 
				To get the states with $\nft=6$ and 8 , 
				we only need to exchange $|0\rangle, |\hat{0}\rangle$ and $|2\rangle, |\hat{2}\rangle$ in states with $\nft=2$ and 0, respectively.
				The states with $\nft=6$ are given by
				\begin{equation}
					\begin{aligned}
						|0222\rangle,\ |\hat{0}\hat{2}\hat{2}\hat{2}\rangle,
						\ |\hat{1}122\rangle,\ |1\hat{1}\hat{2}\hat{2}\rangle,  
						\ |\hat{1}\hat{2}12\rangle\ ,
					\end{aligned}
				\end{equation}
				and cyclic permutated states, whose number is 20. 
				Finally, for states with $\nf=8$ , we have the two states,
				\begin{align}
					|2222\rangle,\ |\hat{2}\hat{2}\hat{2}\hat{2}\rangle\ .
				\end{align}
				Thus, the number of all possible global states is 82.
				Even though we have 82 global states totally, 
				we can deal with them piece by piece because of the fermion number conservation.
				Later, we focus on the set of states with the vanishing net fermion number, which is characterized by $\nft=4$.
				
				\subsection{Actions of the Hamiltonian}\label{23}
				In this section, we introduce the action of the Hamiltonian on the set of states with the vanishing net fermion number, 
				given in Eq.~\refeq{Eq:state4}, which means to calculate their matrix elements in the basis vectors.
				The general action of a gauge-invariant operator on states is derived for the LSH basis in Ref.~\cite{P:2019qdq} and for the angular momentum basis of pure gauge fields in Ref.~\cite{Hayata:2020xxm}. However, it should be noted that there are slight differences in the definitions between these references. We summarize the action of a gauge-invariant operator on states in our notation in Appendix~\ref{A2}.
				
				The electric part Hamiltonian in our model is given as
				\begin{align}
					H_{\rm E}
					&= K\sum_{\mu,\boldsymbol{x}} E^{2}(\boldsymbol{x},\mu)
					\nonumber\\
					&= K\left[ E^2_{{\rm L_1}} + E^2_{{\rm L_2}} + E^2_{{\rm L_3}} + E^2_{{\rm L_4}} \right]\ .
				\end{align}
				It should be noted that $L_i$ does not specify the left electric field here; it simply denotes the label of the link, 
				as shown in the left of Fig.~\ref{Fig:plaquette}.
				Since $E^2_{{\rm L_i}}$ is a Casimir operator, it acts on the local state as
				\begin{equation}
					E_{L_i}^2|s,r_1,r_2,v\rangle=j_i(j_i+1)|s,r_1,r_2,v\rangle\ ,
				\end{equation}
                where $j_i$ can be either $j_a$ or $j_b$ on that link. It is obvious that the global states given in the previous section is 
				the eigenstates of $H_{\rm E}$, the sum of the Casimir operators. 
				The action of $H_{\rm E}$ on global states is determined by the number of hatted labels, such as
				\begin{align}
					&|\hat{1}120\rangle\to\frac{3}{4}|\hat{1}120\rangle\ ,\\ 
					&|\hat{1}\hat{2}10\rangle\to\frac{3}{2}|\hat{1}\hat{2}10\rangle\ ,\\ 
					&|\hat{1}\hat{2}\hat{0}1\rangle\to\frac{9}{4}|\hat{1}\hat{2}\hat{0}1\rangle\ ,\\
					&|\hat{2}\hat{2}\hat{0}\hat{0}\rangle\to3|\hat{2}\hat{2}\hat{0}\hat{0}\rangle\ ,
				\end{align}
				where we omit the coupling constant $K$ for convenient. 
				
				For the magnetic part of the Hamiltonian, we have specific Hamiltonian as
				\begin{align}\label{Hb}
					&H_{\rm B}=-\frac{1}{2K}\sum_{i,j,k,l=\pm} \nonumber\\
					&\times
					\mathcal{L}^{i,j}(a_{1},b_{1})
					\mathcal{L}^{j,k}(a_{2},b_{2})
					\mathcal{L}^{k,l}(a_{3},b_{3})
					\mathcal{L}^{l,i}(a_{4},b_{4})+(\mathrm{h.c.})\ .
				\end{align}
				where the operator $\mathcal{L}$ is the gauge invariant operators~\cite{Raychowdhury:2019iki}.
				The pure boson gauge invariant operator are given as
				\begin{align}
					\mathcal{L}^{++}\left(a, b\right)&=\frac{1}{\sqrt{N_{\mathrm{b}  }+1}}\left(\boldsymbol{a}^{\dagger} \cdot \tilde{\boldsymbol{b}}^{\dagger}\right) \frac{1}{\sqrt{N_{\mathrm{a} }+1}}\ ,\\
					\mathcal{L}^{-+}\left(a, b\right)&=-\frac{1}{\sqrt{N_{\mathrm{b} }+1}}\left(\boldsymbol{a} \cdot \boldsymbol{b}^{\dagger}\right) \frac{1}{\sqrt{N_{\mathrm{a} }+1}}\ ,\\
					\mathcal{L}^{+-}\left(a, b\right)&=\frac{1}{\sqrt{N_{\mathrm{b} }+1}}\left(\boldsymbol{a}^{\dagger} \cdot \boldsymbol{b}\right) \frac{1}{\sqrt{N_{\mathrm{a} }+1}}\ ,\\
					\mathcal{L}^{--}\left(a, b\right)&=\frac{1}{\sqrt{N_{\mathrm{b} }+1}}(\boldsymbol{a} \cdot \tilde{\boldsymbol{b}}) \frac{1}{\sqrt{N_{\mathrm{a} }+1}}\ ,
				\end{align}
				where the $N_\mathrm{a}$ and $N_\mathrm{b}$ denote the number operator of $a$ and $b$ boson respectively. The actions of these operators are given in Appendix~\ref{A2}
				Note that there are only pure boson operators in Eq.~\refeq{Hb}. Therefore, the actions of $H_{\rm B}$ do not change the numbers of fermion at each sites. Besides, the summation in Eq.~\refeq{Hb} indicates that the actions of $H_{\rm B}$ add or subtract one boson at each links. Therefore, in our notation, the actions of $H_{\rm B}$ always add a hat on the states without hat and subtract the hat on the states with a hat.
				Therefore, under the current truncation, $j_{\rm max}=\frac{1}{2}$,
				indeed, we only have one non-vanishing actions of $H_{\rm B}$ on each global state as
				\begin{align}
					&|0220\rangle\to2|\hat{0}\hat{2}\hat{2}\hat{0}\rangle\ ,\\
					&|02\hat{1}1\rangle\to-|\hat{0}\hat{2}1\hat{1}\rangle\ ,\\
					&|\hat{1}120\rangle\to-|1\hat{1}\hat{2}\hat{0}\rangle\ ,\\
					&|\hat{1}\hat{2}10\rangle\to-|12\hat{1}\hat{0}\rangle\ ,\\
					&|\hat{1}1\hat{1}1\rangle\to\frac{1}{2}|1\hat{1}1\hat{1}\rangle\ ,\\
					&|12\hat{1}\hat{0}\rangle\to-|\hat{1}\hat{2}10\rangle\ ,\\
					&|1\hat{1}\hat{2}\hat{0}\rangle\to-|\hat{1}120\rangle\ ,\\
					&|\hat{0}\hat{2}1\hat{1}\rangle\to-|02\hat{1}1\rangle\ ,\\
					&|\hat{0}\hat{2}\hat{2}\hat{0}\rangle\to2|0220\rangle\ ,
				\end{align}
				where we have also ignored the factor $-\frac{1}{2K}$; see also Appendix~\ref{A2} for some details. Here, we present only representative examples, as performing an $n$-fold cyclic permutation of the labels on both sides of the states mentioned above also results in a non-vanishing action of $H_{\rm B}$. It is observed that the magnetic part of the Hamiltonian changes the quantum number of bosons on every site.
				
				For the fermionic part, we first divide the Hamiltonian shown in Eq.~\eqref{Eq:Hk} into two parts as $H_{\rm F}=H_{\rm FK}+H_{\rm FM}$ 
				where $H_{\rm FK}$ and $H_{\rm FM}$ denote the kinetic term and the mass term, respectively. 
				Therefore, in our model, the mass term is given as 
				\begin{equation}
					H_{\rm FM} = -m\psi^\dagger_{\rm A}\psi_{\rm A}
					+m\psi^\dagger_{\rm B}\psi_{\rm B}
					-m\psi^\dagger_{\rm C}\psi_{\rm C}
					+m\psi^\dagger_{\rm D}\psi_{\rm D}\ .
				\end{equation}
				The minus sign reflects the fact that there is anti-particle at the sites, A and C. 
				It is not difficult to write down the non-vanishing action of $H_{\rm FM}$ as  
				\begin{align}
					&|2020\rangle\to-4|2020\rangle\ ,\\ 
					&|0202\rangle\to4|0202\rangle\ ,\\ 
					&|\hat{1}120\rangle\to-2|\hat{1}120\rangle\ ,\\ 
					&|\hat{0}1\hat{1}\hat{2}\rangle\to2|\hat{0}1\hat{1}\hat{2}\rangle\ ,\\ 
					&|102\hat{1}\rangle\to-2|102\hat{1}\rangle\ ,\\ 
					&|\hat{1}102\rangle\to2|\hat{1}102\rangle\ , 
				\end{align}
				where we omit the mass factor $m$. 
				As with listing the non-vanishing actions of $H_{\rm B}$, 
				we show only representative examples here, as performing a $2n$-fold cyclic permutation on the labels on both sides of the states listed above also results in a non-vanishing action of $H_{\rm FM}$.
				
				Finally, for the kinetic term in our model, we can rewrite it using gauge invariant operators as
				\begin{equation}
					\begin{aligned}
						&H_{\rm FK}\\
						=&\left[\mathcal{S}^{++}(\psi_{\rm A}, b_1)\mathcal{S}^{+-}(a_2,  \psi_{\rm B})
						+\mathcal{S}^{+-}(b_1, \psi_{\rm A})\mathcal{S}^{++}(\psi_{\rm B},a_2)  \right]\\
						+&\left[\mathcal{S}^{++}(\psi_{\rm B}, b_2)\mathcal{S}^{+-}(a_3,  \psi_{\rm C})
						+\mathcal{S}^{+-}(b_2, \psi_{\rm B})\mathcal{S}^{++}(\psi_{\rm C},a_3)  \right]\\
						+&\left[\mathcal{S}^{++}(\psi_{\rm C}, b_3)\mathcal{S}^{+-}(a_4,  \psi_{\rm D})
						+\mathcal{S}^{+-}(b_3, \psi_{\rm C})\mathcal{S}^{++}(\psi_{\rm D},a_4)  \right]\\
						+&\left[\mathcal{S}^{+-}(a_1, \psi_{\rm A})\mathcal{S}^{++}(\psi_{\rm D},b_4)  
						+\mathcal{S}^{++}(\psi_{\rm A}, a_1)\mathcal{S}^{+-}(b_4,  \psi_{\rm D})\right]\\
						+&\ (\mathrm{h.c.})\ ,\label{Eq:hfk}
					\end{aligned} 
				\end{equation}
				which is a summation of interaction terms acting on two neighboring sites; see Ref.~\cite{Raychowdhury:2019iki} for details.
				The gauge-invariant operator $\mathcal{S}$ containing one boson and one fermion operators is given as follows:
				\begin{align}
					\mathcal{S}^{++}\left(\psi, b\right)&=\frac{1}{\sqrt{N_{\mathrm{b}  }+1}}\left(\boldsymbol{\psi}^{\dagger} \cdot \tilde{\boldsymbol{b}}^{\dagger}\right)\ ,\\
					\mathcal{S}^{++}\left(\psi, a\right)&=\frac{1}{\sqrt{N_{\mathrm{a}  }+1}}\left(\boldsymbol{\psi}^{\dagger} \cdot \tilde{\boldsymbol{a}}^{\dagger}\right)\ ,\\
					\mathcal{S}^{+-}\left(b, \psi\right)&=\left(\boldsymbol{b}^{\dagger} \cdot \boldsymbol{\psi}\right) \frac{1}{\sqrt{N_{\mathrm{b} }+1}}\ ,\\
					\mathcal{S}^{+-}\left(a, \psi\right)&=\left(\boldsymbol{a}^{\dagger} \cdot \boldsymbol{\psi}\right) \frac{1}{\sqrt{N_{\mathrm{a} }+1}}\ ,
				\end{align} 
				whose actions are also shown in Appendix~\ref{A2}. In Eq.~\refeq{Eq:hfk}, the first through fourth brackets act in the same way on the sites at both ends of links 1 through 4, respectively. 
				The Hermitian conjugate terms correspond to actions that reverse the processes applied by these four brackets.
				Thus, we only need to determine all possible actions on both ends of link 1 to fully understand the action of $H_{\rm FK}$. 
				To do this, we first consider how the gauge-invariant operators in the first bracket acts on the states located at sites A and B.
				Then, all the non-vanishing actions of the single gauge-invariant operator included in the first bracket are given as follows:
				\begin{align}
					&\mathcal{S}^{++}(\psi_{\rm A}, b_1)|0n_{\rm B}\rangle      =-                  |\hat{1}n_{\rm B}\rangle\ ,
					\\&\mathcal{S}^{++}(\psi_{\rm A}, b_1)|1n_{\rm B}\rangle      =-\frac{\sqrt{2}}{2}|\hat{2}n_{\rm B}\rangle \ ,
					\\&\mathcal{S}^{++}(\psi_{\rm B}, a_2)|n_{\rm A}0\rangle      =-                  (-1)^{n_{\rm A}} |n_{\rm A}1\rangle      \ ,
					\\&\mathcal{S}^{++}(\psi_{\rm B}, a_2)|n_{\rm A}\hat{1}\rangle= (-1)^{n_{\rm A}} \frac{\sqrt{2}}{2}|n_{\rm A}\hat{2}\rangle\ ,
					\\&\mathcal{S}^{+-}(b_1, \psi_{\rm A})|2n_{\rm B}\rangle      = \sqrt{2}          |\hat{1}n_{\rm B}\rangle\ ,
					\\&\mathcal{S}^{+-}(b_1, \psi_{\rm A})|1n_{\rm B}\rangle      =-                  |\hat{0}n_{\rm B}\rangle\ ,
					\\&\mathcal{S}^{+-}(a_2, \psi_{\rm B})|n_{\rm A}2\rangle      = (-1)^{n_{\rm A}} \sqrt{2}          |n_{\rm A}1\rangle      \ ,
					\\&\mathcal{S}^{+-}(a_2, \psi_{\rm B})|n_{\rm A}\hat{1}\rangle= (-1)^{n_{\rm A}}                  |n_{\rm A}\hat{0}\rangle\ ,
				\end{align}
				where $n_{\rm A}$ and $n_{\rm B}$ are the labels of the local states at the sites, $A$ and $B$, respectively.
				By combining these actions, we can obtain all the non-vanishing actions of first bracket terms as
				\begin{align}
					&|20\rangle\to-\sqrt{2}|\hat{1}1\rangle
					\ ,\\&|02\rangle\to-\sqrt{2}|\hat{1}1\rangle
					\ ,\\&|1\hat{1}\rangle\to\frac{\sqrt{2}}{2}(|\hat{2}\hat{0}\rangle+|\hat{0}\hat{2}\rangle)
					\ ,\\&|0\hat{1}\rangle\to-|\hat{1}\hat{0}\rangle   
					\ ,\\&|10\rangle\to-|\hat{0}1\rangle
					\ ,\\&|2\hat{1}\rangle\to|\hat{1}\hat{2}\rangle
					\ ,\\&|12\rangle\to|\hat{2}1\rangle\ .
				\end{align}
				To fully describe the actions of $H_{\rm FK}$, we only need to apply the same transformation rule to every pair of neighboring sites.
				Therefore, we can obtain them as 
				\begin{small}
					\begin{align}
						&|0220\rangle\to-\sqrt{2}|\hat{1}120\rangle-\sqrt{2}|02\hat{1}1\rangle\ ,\\
						&|2020\rangle\to-\sqrt{2}|\hat{1}120\rangle-\sqrt{2}|20\hat{1}1\rangle-\sqrt{2}|2\hat{1}10\rangle-\sqrt{2}|102\hat{1}\rangle
						\ ,\\ &|\hat{1}120\rangle\to|\hat{1}\hat{2}10\rangle-\sqrt{2}|\hat{1}1\hat{1}1\rangle-|\hat{0}12\hat{1}\rangle\ ,\\ 
						&|02\hat{1}1\rangle\to|0\hat{1}\hat{2}1\rangle-\sqrt{2}|\hat{1}1\hat{1}1\rangle-|12\hat{1}\hat{0}\rangle
						\ ,\\ &|\hat{1}\hat{2}10\rangle\to-|\hat{1}\hat{2}\hat{0}1\rangle-|\hat{0}\hat{2}1\hat{1}\rangle\ ,\\
						&|\hat{0}12\hat{1}\rangle\to-|\hat{0}1\hat{1}\hat{2}\rangle+|\hat{0}\hat{2}1\hat{1}\rangle\ ,\\
						&|\hat{1}1\hat{1}1\rangle\to\frac{\sqrt{2}}{2}(|\hat{1}\hat{2}\hat{0}1\rangle+|\hat{1}\hat{0}\hat{2}1\rangle+|\hat{0}1\hat{1}\hat{2}\rangle+|\hat{2}1\hat{1}\hat{0}\rangle)  
						\ ,\\&|\hat{1}\hat{2}\hat{0}1\rangle\to\frac{\sqrt{2}}{2}|\hat{2}\hat{2}\hat{0}\hat{0}\rangle+\frac{\sqrt{2}}{2}|\hat{0}\hat{2}\hat{0}\hat{2}\rangle\ ,\\
						&|\hat{0}\hat{2}1\hat{1}\rangle\to\frac{\sqrt{2}}{2}|\hat{0}\hat{2}\hat{2}\hat{0}\rangle+\frac{\sqrt{2}}{2}|\hat{0}\hat{2}\hat{0}\hat{2}\rangle\ ,
					\end{align}
				\end{small}and their inverse process is described by the Hermitian conjugate terms.
				Here we show only representative examples, 
				as performing an $n$-times cyclic permutation on the labels on both sides of the states listed above also results in a non-vanishing action of $H_{\rm FK}$.
				Note that the actions always add or remove a hat on the total states. 
				
				Now, all actions of the total Hamiltonian are provided as matrix elements in the specific basis. 
				For quantum computation purposes, we need to translate these matrix elements into the spin basis.
				
				\section{Quantum Simulation}\label{Sec3}
				
				\subsection{Mapping to spin system}
				
				To perform quantum simulations, we need to map our model into a spin system. Then, we can write down the Hamiltonian matrix using the identity and Pauli matrices, which together can represent any general $2\times2$ Hermitian matrix. Since the number of the global states we consider is 38, we need 6 qubits at least to represent them. Note that there are specific rules for the number of the hats of the global states that are changed when the Hamiltonian acts on them, as discussed in Sec.~\ref{23}. This suggests that dividing the global states into different groups is convenient.
				Therefore, we classify the global states into 5 groups based on the number of hats.  We use the first 3 qubits to represent the groups as
					\begin{equation}\label{gmapping}
						\begin{aligned}
							\mathrm{group~A}:\qquad& |\uparrow\rangle|\uparrow\rangle|\uparrow\rangle \ ,\\
							\mathrm{group~B}:\qquad& |\uparrow\rangle|\uparrow\rangle|\downarrow\rangle \ ,\\
							\mathrm{group~C_{1}}:\qquad& |\uparrow\rangle|\downarrow\rangle|\uparrow\rangle \ ,\\
							\mathrm{group~C_{2}}:\qquad & |\uparrow\rangle|\downarrow\rangle|\downarrow\rangle \ ,\\
							\mathrm{group~D}:  \qquad   & |\downarrow\rangle|\uparrow\rangle|\uparrow\rangle \ ,\\
							\mathrm{group~E}:  \qquad   & |\downarrow\rangle|\uparrow\rangle|\downarrow\rangle \ ,
						\end{aligned}
					\end{equation}
				where the groups A, B, C, D, and E correspond to the global states with 0, 1, 2, 3, and 4 hats, respectively. The group C has 10 patterns, so we divide it into the group $\mathrm{C_1}$ and $\mathrm{C_2}$: states where hats are adjacent are classified as the group $\mathrm{C_1}$, while all other states are classified as the group $\mathrm{C_2}$. Then we use the left 3 qubits to specify the state in each group, as shown in Appendix~\ref{B1}.
				Based on this mapping, regardless of which term in the Hamiltonian acts, all global states within one group are transformed into states within a single specific group, which may be the same as or different from the original group, which is convenient for figuring out the Hamiltonian matrix.
				
				Now, we can write down the Hamitonian matrix with Pauli matrices $X=\sigma^1$, $Y=\sigma^2$, $Z=\sigma^3$ and identity matrix $I$. Since there are numerous terms, we just take an example to show the steps and give the full results in Appendix~\ref{B2}. Let's focus on the action, $|\hat{1}120\rangle\to-\sqrt{2}|\hat{1}1\hat{1}1\rangle$, which is the action from the group B to the group $\mathrm{C_2}$. Thus, we can directly find the matrix for the first 3 qubits based on the mapping rule for the groups~\eqref{gmapping} as
				\begin{align}
					\left[\begin{array}{c c}
						{{1}}&{{0}}\\
						{{0}}&{{0}}
					\end{array}\right] 
					\otimes
					\left[\begin{array}{c c}
						{{0}}&{{0}}\\
						{{1}}&{{0}}
					\end{array}\right] 
					\otimes
					\left[\begin{array}{c c}
						{{0}}&{{0}}\\
						{{0}}&{{1}}
					\end{array}\right]\ . 
				\end{align}
				Then, according to the mapping rule (\ref{B1}), we can write the matrix for the last 3 qubits as
				\begin{align}
					-\sqrt{2}
					\left[\begin{array}{c c}
						{{1}}&{{1}}\\
						{{0}}&{{0}}
					\end{array}\right] 
					\otimes
					\left[\begin{array}{c c}
						{{1}}&{{1}}\\
						{{0}}&{{0}}
					\end{array}\right]
					\otimes
					\left[\begin{array}{c c}
						{{1}}&{{0}}\\
						{{0}}&{{1}}
					\end{array}\right]\ .
				\end{align}
				Using Pauli matrices, we can represent the term and its transpose as
				\begin{equation}
					\begin{aligned}
						&-\frac{\sqrt{2}}{16}
						(I+Z)\{X(I-Z)[(I+Z+X)(I+Z+X)-YY]\\
						&\quad+Y(I-Z)[(I+Z+X)Y+Y(I+Z+X)]\}I\ ,
					\end{aligned}
				\end{equation}
				where we ignore the symbol $\otimes$ for simplicity. 
				We show all results in Appendix~\ref{B2}. 
				Based on these Hamiltonian matrices, we can perform quantum simulations which are shown in the next section. 
				
				\subsection{Real-time simulation}
				Here we introduce our quantum simulation algorithm. The real-time evolution of a quantum system is described by the Schrodinger equation as 
				\begin{equation}
					i\partial_t |\psi(t)\rangle=\hat{H}|\psi(t)\rangle\ .
				\end{equation}
				Since the Hamiltonian in our model is time-independent, the evolution of the quantum state can be described as
				\begin{equation}
					|\psi(t)\rangle=e^{-i \hat{H}t}|\psi(0)\rangle\ ,
				\end{equation}
				where $e^{-i \hat{H}t}$ denotes the real-time evolution operator.
				Note that, in quantum simulation, the quantum gate for a real-time evolution operator can be very difficult to construct. Therefore, we need to construct the quantum gate approximately using the so-called the Suzuki-Trotter decomposition formula~\cite{trotter1959product,suzuki1976generalized} which is
				\begin{equation}
					e^{i(H_1+H_2)\delta t}=e^{iH_1 \delta t}e^{iH_2 \delta t}+O(\delta t^2)\ ,
					\label{eq:std}
				\end{equation}
				where $H_1$ and $H_2$ are two non-commuting operators. Then, for a long time quantum simulation, we can divide the time evolution operator into many small steps as
				\begin{equation}
					\lim _{n \rightarrow \infty}\left(e^{i H_1 \frac{t}{n}} e^{i H_2 \frac{t}{n}}\right)^{n}=e^{i(H_1+H_2) t}\ ,
				\end{equation}
				which can be generalized to the situation with more non-commuting operators. Since we have represented the total Hamiltonian with Pauli matrices, the quantum gate $e^{iH_j t}$ can be exactly constructed where $H_j$ can be any single term in the total Hamiltonian.

				Up to this point, we provide the mapping of the lattice gauge theory with the staggered fermion to the spin system. Through such a mapping, the ultimate goal is to perform quantum computations of the gauge theory. However, the current mapping involves numerous non-local operations, which are significantly affected by noise on existing quantum computers, presenting challenges for practical implementation. For this reason, we performed the calculations using the simulator, Qiskit provided by IBM~\cite{Javadi-Abhari:2024kbf}. Even though calculations on quantum computers are currently difficult, clarifying the correspondence between the specific gauge theory and a spin system is an advancement both theoretically and computationally. Furthermore, the proposed mapping is expected to serve as an important building block for developing more efficient mappings and algorithms in the future.

				Later in this section, in order to investigate the thermalization of non-abelian gauge theory with fermions and the process of fermion production, we calculate the evolution of the entanglement and the pair production. 
				
				\subsubsection{Entanglement of the system}

                \begin{figure*}[htp]
                
                \begin{minipage}{0.3\textwidth}
						\includegraphics[width=220px]{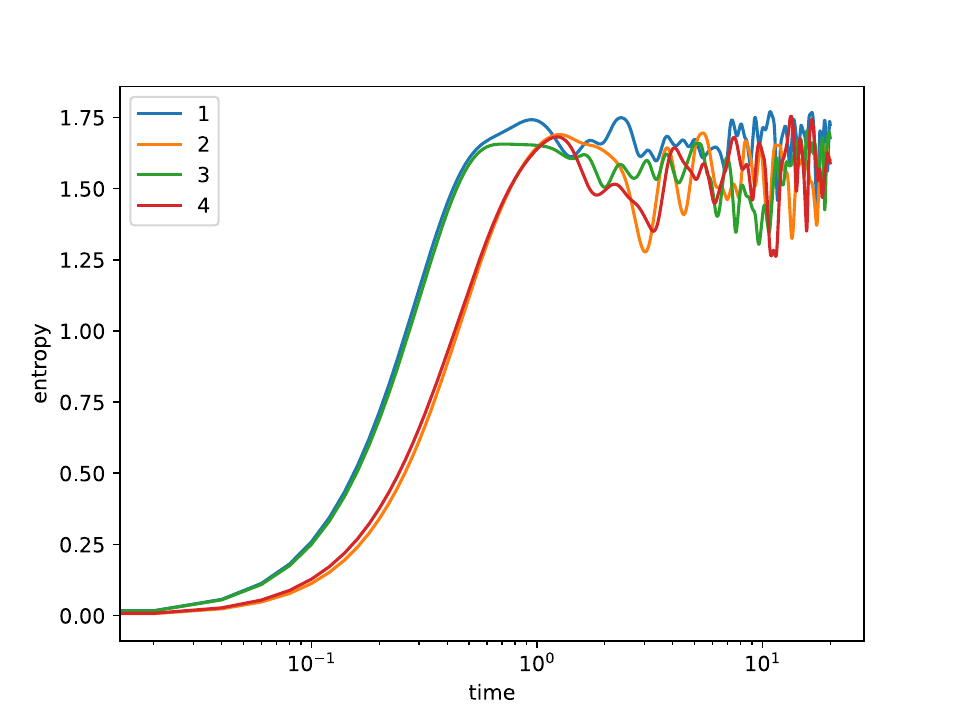}
					\end{minipage}
					\hspace{7.5em}
				\begin{minipage}{0.4\textwidth}
						\includegraphics[width=220px]{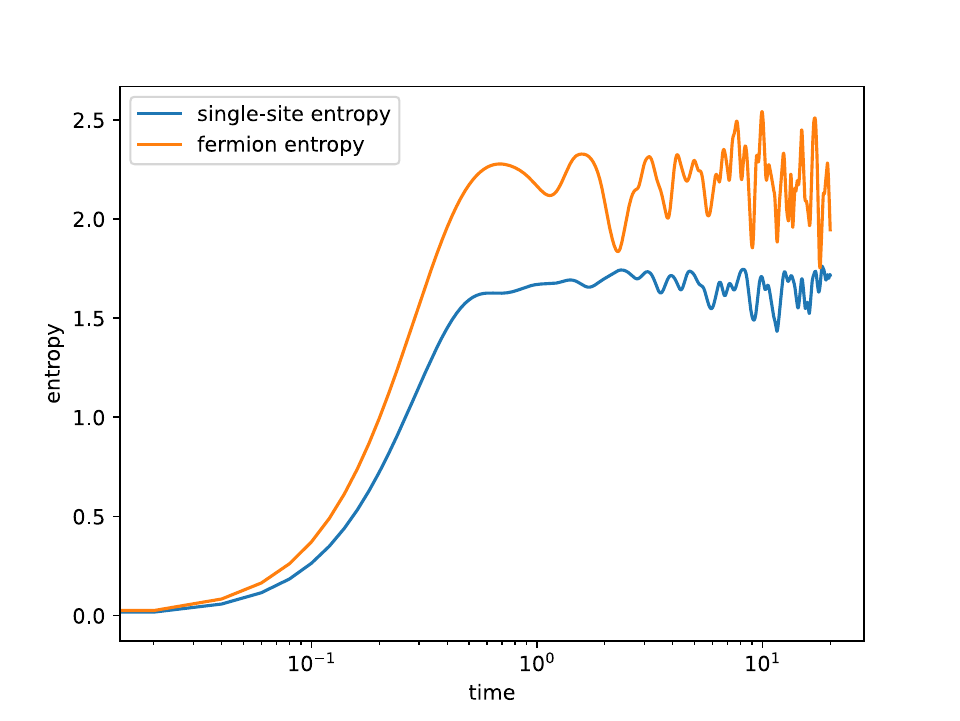}
				\end{minipage}
                 \caption{
                         Left:  Time evolution of the single-site entropy with different initial conditions. Labels "1,2,3,4" denote the time evolution of $|1\hat{1}1\hat{1}\rangle,|\hat{1}\hat{2}10\rangle,|\hat{1}120\rangle,|\hat{1}\hat{2}\hat{0}1\rangle$, respectively.
                         Right:Time evolution of both two kinds of entropy with initial state         
                        $|\hat{1}201\rangle$. In both figures, we set the parameters as $K=1,m=1$ which make the eigenvalue of $H_E,H_B,H_{FK},H_{FM}$ be close. This ensures the balance among the four parts of the Hamiltonian. 
                 }
                \label{Fig:entropy}
                \end{figure*}
				To measure the thermalization of non-Abelian gauge theory with fermions, we first introduce two kinds of entanglement entropy. The entanglement entropy quantifies the mixedness of a subsystem of interest, arising from the entanglement between the subsystem and the rest of the system.
				The density matrix of the whole system is defined as
				\begin{equation}
					\hat{\rho}(t)=|\psi(t)\rangle\langle\psi(t)|\ .
				\end{equation}
				Focusing on the entanglement between the local state at the single site $A$ and states on the other sites, we define the reduced density matrix by tracing out the Hilbert space at sites B,C, and D as
				\begin{equation}
					\hat{\rho}_{\text{A}}(t)=\sum_{i \in \text{all basis states}}\langle\psi^i_{\text{BCD}}| \hat{\rho}(t)|\psi^i_{\text{BCD}}\rangle\ ,
				\end{equation}
				where A, B, C, and D denote the labels of sites shown in Fig.~\ref{Fig:plaquette} and $|\psi^i_{\text{BCD}}\rangle$ denotes a basis vector for the Hilbert space at the sites B, C, and D: for example $|0_{\text{B}} 2_{\text{C}} 0_{\text{D}}\rangle$. Note that the local state on the point A can be uniquely determined by the states on the sites B,C, and D because of the constraint (\ref{connect}). Therefore, the reduced density matrix $\hat{\rho}_{\text{A}}(t)$ has only diagonal elements which denote the probabilities of being in the corresponding local states. Then, the entanglement entropy is defined as 
				\begin{equation}
					S_{\text{A}}(t)=-{\rm Tr}\left[\hat{\rho}_{\text{A}}(t)\ln{\hat{\rho}_{\text{A}}(t)}\right]\ .
				\end{equation}
				Later we refer to $S_{\text{A}}(t)$ as the single-site entanglement entropy.
				
				In addition to the entanglement entropy for a single lattice site, we also consider the entanglement entropy defined from the density matrix obtained by tracing out the boson Hilbert space. This entropy corresponds to the entropy carried by fermions and is more directly related to the entropy estimated from, for example, the hadron multiplicities observed in relativistic heavy-ion collision experiments. The reduced density matrix is defined as
				\begin{equation}
					\hat{\rho}_{\text{F}}(t)=\sum_{i\in\text{all basis states}}\langle\psi^i_{\text{boson}}| \hat{\rho}(t)|\psi^i_{\text{boson}}\rangle\ ,
				\end{equation}
				where $|\psi^i_{\text{boson}}\rangle$ denotes the state for Schwinger bosons. Note that $|\psi_{\text{boson}}^i\rangle$ can be identified as the total state without fermion number on the sites. 
				Finally, we have another entangle entropy as
				\begin{equation}
					S_{\text{F}}(t)=-{\rm Tr}\left[\hat{\rho}_{\text{F}}(t)\ln{\hat{\rho}_{\text{F}}(t)}\right]\ .
				\end{equation}
				Later we refer to $S_{\text{F}}(t)$ as the fermion entropy.
				
				We show the time evolution of the single-site entanglement entropy with four different initial conditions, 
				$|1\hat{1}1\hat{1}\rangle,|\hat{1}\hat{2}10\rangle,|\hat{1}120\rangle,|\hat{1}\hat{2}\hat{0}1\rangle$ at $K=1$ and $m=1$, in the left of Fig.~\ref{Fig:entropy}.
		
				As expected, the entropies are found to be initially 0, increase rapidly at first, and then fluctuate around some constant. 
			
				This finding indicates that, among the initial conditions we test, the system always reaches some kind of ``thermalized" state.
				The observation that thermal equilibration is implied, even though the system size is significantly small, is consistent with the results in Ref.~\cite{Hayata:2020xxm}, which reports numerical simulations of the pure SU(2) gauge theory on a small lattice.
				We should note that the fluctuations are visually large. This may reflect the fact that the Hilbert space we are considering is small.

				We also show the real-time evolution of both two kinds of entanglement entropy on the right of Fig.~\ref{Fig:entropy}. We can clearly find that these two kinds of entropy have the same behavior as the time evolves. Since the dimension of Hilbert space for a single site is 6, the single-site entropy mathematically has a maximum value which is $\ln 6 \, (\approx1.79)$. For the same reason, fermion entropy also has a maximum value which is $\ln 16 \, (\approx2.77)$. Therefore, these two entropies reach high values, exceeding 80\% of their respective maximum values.
				This indicates that, in the current setup, the reduced density matrices for the single-site and fermion subsystems spread widely within their respective Hilbert spaces over a long time.

				Note that in our Hamiltonian, we have two parameters: the fermion mass $m$ and the coupling constant $K$.
				Regarding the fermion mass $m$, we present the result in the next section that, as $m$ increases, fermion excitation is suppressed, and accordingly, the fermion entropy is also suppressed.
				Regarding the coupling constant $K$, throughout this paper, we fix 
				$K=1$ for practical reasons. 
				Since the electric part of the Hamiltonian is proportional to the coupling constant, $H_{\rm E} \propto K$, and the magnetic part is 
				proportional to the inverse of the coupling constant, $H_{\rm B} \propto K^{-1}$, both smaller and larger values of $K$ lead to an enhancement of the Hamiltonian. 
				As a result, numerical simulations become more challenging: a smaller $\delta t$ is required to maintain the validity of the Suzuki-Trotter decomposition \eqref{eq:std}.
				The investigation of $K$-dependence in this model is beyond the scope of this work.

				\subsubsection{Pair production}
				\begin{figure*}[tp]
					\begin{minipage}{0.3\textwidth}
						\includegraphics[width=220px]{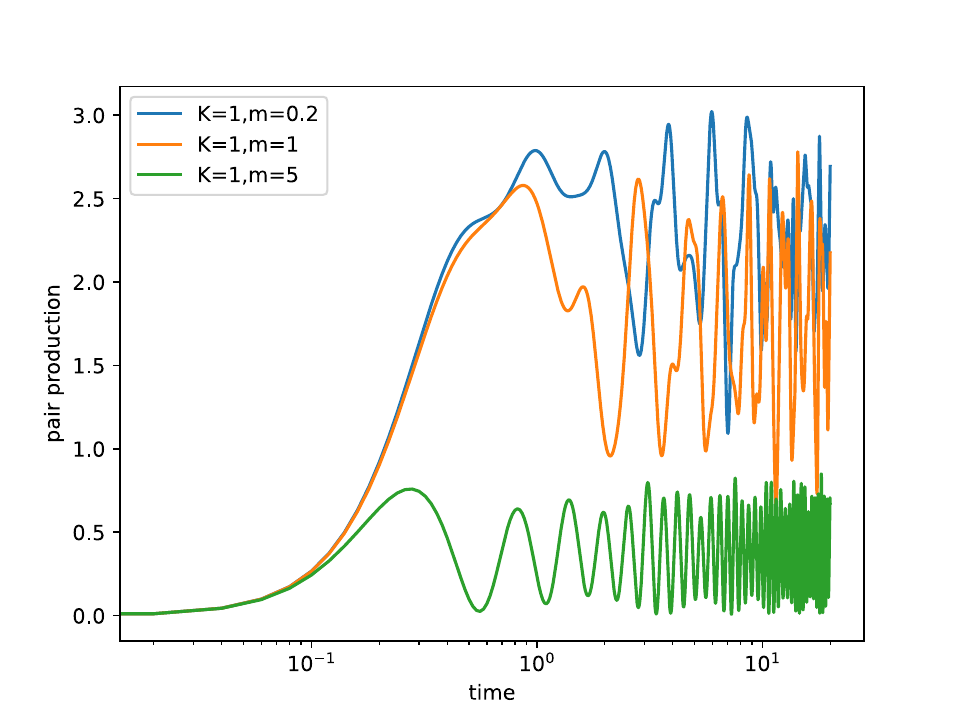}
					\end{minipage}
					\hspace{7.5em}
					\begin{minipage}{0.4\textwidth}
						\includegraphics[width=220px]{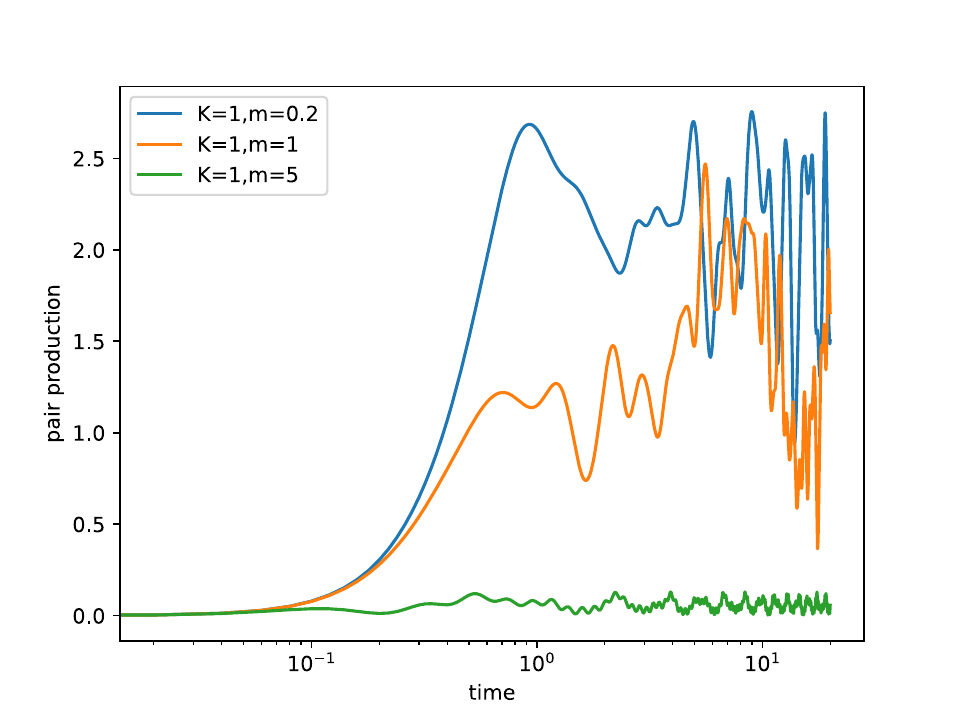}
					\end{minipage}
					\caption{
						Left:  
						Time evolution of the pair productions with initial state $|2020\rangle$ and different fermion mass $m$. 
						Right:
						Time evolution of the pair productions with initial state $|\hat{2}\hat{0}\hat{2}\hat{0}\rangle$ and different fermion mass $m$. In both figures, we set coupling constant as $K=1$.
					}
					\label{Fig:pair production}
				\end{figure*}
				
				\begin{figure*}[tp]
					\begin{minipage}{0.3\textwidth}
						\includegraphics[width=220px]{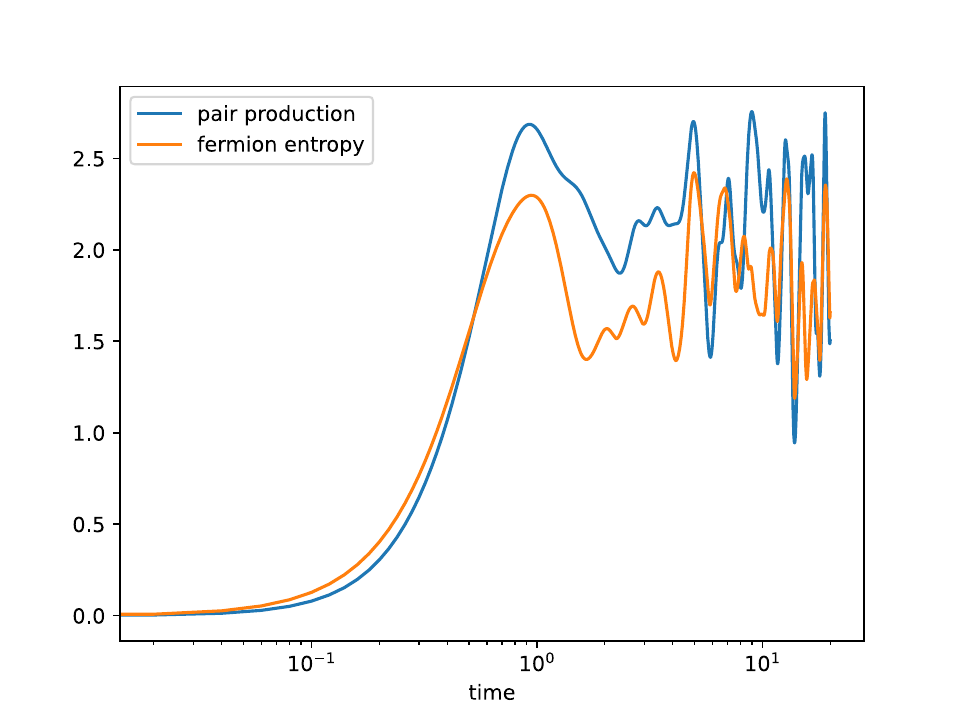}
					\end{minipage}
					\hspace{7.5em}
					\begin{minipage}{0.4\textwidth}
						\includegraphics[width=220px]{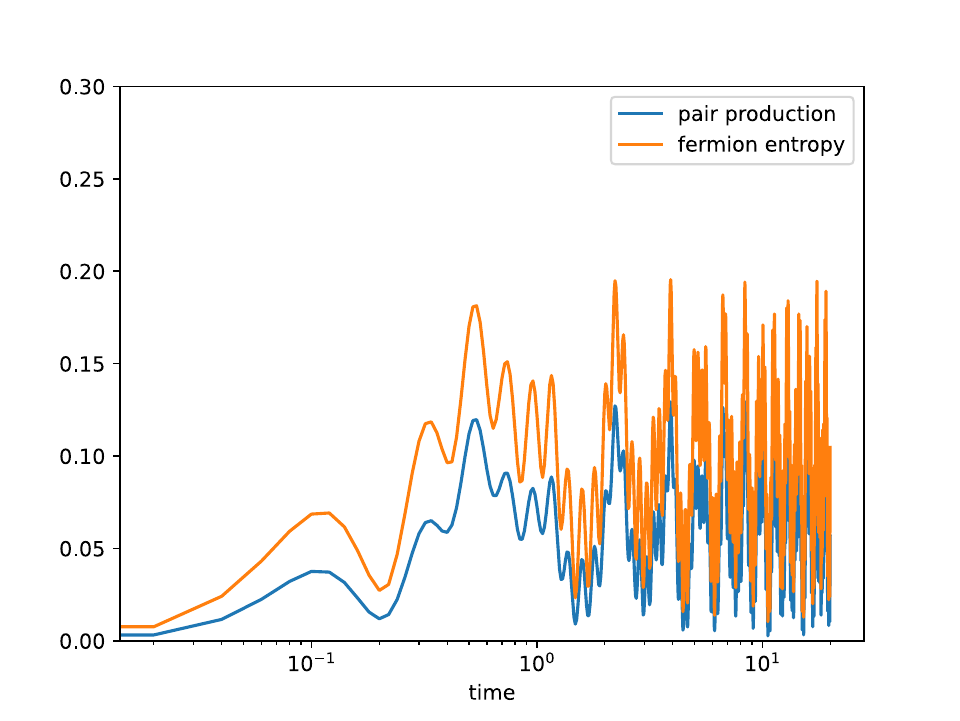}
					\end{minipage}
					\caption{
						Left:  
						Time evolution of pair production in comparison of the fermion entropy with $K=1,m=0.2$.
						Right:
						Time evolution of pair production in comparison of the fermion entropy with $K=1,m=5$. In both figures, the initial states are $|\hat{2}\hat{0}\hat{2}\hat{0}\rangle$.
					}
					\label{Fig:comparision}
				\end{figure*}
				In our simulation, we choose the global states which have vanishing fermion number. 
				Because of the fermion number conservation,  fermions and anti-fermions are always excited in pairs. 
				Thus, we refer to the fermion number, associated with the sites at point B and D, as the pair production,
				\begin{equation}
					N_{\rm pair} \equiv 4-N_{{\rm f},A}-N_{{\rm f},C}=N_{{\rm f},B}+N_{{\rm f},D}\ .
				\end{equation}
			
                We show the time evolution of the pair productions in Fig.~\ref{Fig:pair production} at $1$ and $m=0.2, 1$ and $5$ with initial conditions, $|2020\rangle$ and $|\hat{2}\hat{0}\hat{2}\hat{0}\rangle$, which are the only states with vanishing pair production, $N_{\rm pair}=0$.  
				 
				In both figures, as expected, when the mass is not too large which means the evolutions for $m=1$ and $m=0.2$, the pair productions increase rapidly at first and then fluctuate around some constants.
				Besides, when the mass is large enough, it should be too difficult to produce fermion pairs. Therefore, in both figures, the pair productions denoted as $m=5$ finally fluctuate at some low levels. We can also clearly see the mass dependence of the pair productions. Both the maximum value and the average value of the pair production decrease as the mass become larger.
				Note that because of the anti-commutation relation of fermion, the pair production is limited by the size of lattice. Therefore, when both two evolutions have small enough masses, the distinctions may be not clear even though the masses are significantly different which is shown in the left of Fig.~\ref{Fig:pair production}.
				The fluctuations are also visually large which may have the same reason with the evolutions of entropies.
				After all, these results still satisfy the physical intuition. 
				
				To study the relation between pair production and fermion entropy, we also show the comparison of their time evolutions in Fig.~\ref{Fig:comparision}. 
				We have already known that the fermion mass can have a great influence on the pair production. Therefore, we show the time evolutions with small mass and large mass on the left and right of Fig.~\ref{Fig:comparision}, respectively. These two results approximately represent the situations where fermions are produced and where fermions are not produced.
				We can clearly see that the fermion entropy is increasing when the fermions being produced and fluctuating when the pair production fluctuating.  Besides, if there is hardly any pair production, the fermion entropy will also be at low level.
				The coincidence between the behavior of the fermion number and the entropy in the non-equilibrium evolution is non-trivial while, in a thermal equilibrium state, both the entropy and the particle number are considered extensive quantities and proportional to each other.
				This coincidence suggests that, even in non-equilibrium, there may be cases where the entropy can be estimated from the fermion number.

				\section{Summary and Outlook}\label{Sec4}
				We have simulated the real-time evolution of SU(2) gauge theory with fermions on a small two-dimensional lattice using the loop-string-hadron (LSH) formulation. The LSH formulation provides a gauge-invariant Hilbert space for gauge theories with fermions and, moreover, allows us to truncate the Hilbert space while preserving gauge invariance.
				We have mapped the formulated model to a spin system, laying the groundwork for future quantum simulations on actual quantum computers. However, we note that the mapping we have introduced involves many nonlocal interactions, which necessitate further algorithmic improvements for practical quantum computations. Then, using a quantum computer emulator, we have tracked the real-time quantum evolution of the system starting from nonequilibrium states and have studied the entanglement entropy and fermion production.
				
				We have investigated the thermalization of the system through the entanglement entropy of a single lattice site. The results show that entropy initially grows significantly and then oscillates around a certain value. This behavior indicates that the entanglement of a single site with other lattice sites increases as it interacts with them, suggesting the thermalization of the system. Even in a small lattice, the system exhibits thermal equilibrium-like behavior, which is consistent with previous studies on pure SU(2) Yang-Mills theory~\cite{Hayata:2020xxm}. Physically, this suggests that the Hilbert space is already dense enough for quantum chaos to emerge.
				
				Furthermore, to explore the relationship between the entropy of fermions and the fermion particle number, we have calculated the entanglement entropy of the density matrix obtained by performing a partial trace over the bosonic Hilbert space, as well as the evolution of the fermion number from an initial nonequilibrium state with a vanishing fermion number. Our results show that, within the range of initial states we have simulated, both quantities exhibit similar behavior. While entropy is known to be proportional to the particle number in equilibrium, the relationship between them in nonequilibrium states remains largely unexplored. This finding suggests that even in far-from-equilibrium systems, there exists a close relationship between the entropy and the particle number production of fermions.
				
				In the future, to achieve our objective of simulating real QCD, we need to improve our model step by step.
				In particular, to investigate more aspects of SU(2) Yang-Mills theory, we need to increase the lattice size and consider higher-energy states, which may provide a testing ground for thermalization.
				Furthermore, we would like to consider a more realistic scenario??namely, SU(3) Yang-Mills theory with fermions. Once we achieve this objective, we hope to gain deeper insights into the physics of QCD.

				\section*{Acknowledgments}
				This work is supported in part by Grants-in-Aid for Scientific Research from Japan Society for the Promotion of Science (JSPS) KAKENHI (Grant No. JP22H05112), by Natural Science Foundation of Shanghai through Grant No. 23JC1400200, by National Natural Science Foundation of China through Grants No. 12225502, No. 12147101, and by National Key Research and Development Program of China through Grant No. 2022YFA1604900.

				\appendix

				\section{Action of gauge invariant operator}\label{A2}
				In this section, we summarize the actions of all necessary gauge invariant operators. For pure boson type operators which appear in the magnetic part of Hamiltonian, we have 
				\begin{align}
					\mathcal{L}^{++}\left(a, b\right)&=\frac{1}{\sqrt{N_{\mathrm{b}  }+1}}\left(\boldsymbol{a}^{\dagger} \cdot \tilde{\boldsymbol{b}}^{\dagger}\right) \frac{1}{\sqrt{N_{\mathrm{a} }+1}}\ ,\\
					\mathcal{L}^{-+}\left(a, b\right)&=-\frac{1}{\sqrt{N_{\mathrm{b} }+1}}\left(\boldsymbol{a} \cdot \boldsymbol{b}^{\dagger}\right) \frac{1}{\sqrt{N_{\mathrm{a} }+1}}\ ,\\
					\mathcal{L}^{+-}\left(a, b\right)&=\frac{1}{\sqrt{N_{\mathrm{b} }+1}}\left(\boldsymbol{a}^{\dagger} \cdot \boldsymbol{b}\right) \frac{1}{\sqrt{N_{\mathrm{a} }+1}}\ ,\\
					\mathcal{L}^{--}\left(a, b\right)&=\frac{1}{\sqrt{N_{\mathrm{b} }+1}}(\boldsymbol{a} \cdot \tilde{\boldsymbol{b}}) \frac{1}{\sqrt{N_{\mathrm{a} }+1}}\ ,
				\end{align}
				which have the same form with formula in Ref.~\cite{Hayata:2020xxm}. Besides, in the fermionic part of Hamiltonian, we have the operators including fermions as
				\begin{equation}\label{mixed}
					\begin{aligned}
						\mathcal{S}^{++}\left(\psi, b\right)&=\frac{1}{\sqrt{N_{\mathrm{b}  }+1}}\left(\boldsymbol{\psi}^{\dagger} \cdot \tilde{\boldsymbol{b}}^{\dagger}\right) \\
						\mathcal{S}^{++}\left(\psi, a\right)&=\frac{1}{\sqrt{N_{\mathrm{a}  }+1}}\left(\boldsymbol{\psi}^{\dagger} \cdot \tilde{\boldsymbol{a}}^{\dagger}\right)\\
						\mathcal{S}^{+-}\left(b, \psi\right)&=\left(\boldsymbol{b}^{\dagger} \cdot \boldsymbol{\psi}\right) \frac{1}{\sqrt{N_{\mathrm{b} }+1}}\\
						\mathcal{S}^{+-}\left(a, \psi\right)&=\left(\boldsymbol{a}^{\dagger} \cdot \boldsymbol{\psi}\right) \frac{1}{\sqrt{N_{\mathrm{a} }+1}}.
					\end{aligned}
				\end{equation}
				Note that, because of the anti-commutation relation of fermions and the order of gauge invariant operators when defining physical states, it is necessary to distinguish bosons $a,b$ in Eq.~\eqref{mixed}. Besides, all $N_a,N_b$ should be treated as operators. Before we introduce the details for calculating the actions, we show the relation between LSH basis and angular momentum basis again as
				\begin{equation}
					j_\mathrm{b}=\frac{1}{2}\left(s+r_{1}\right), \quad j_\mathrm{a}=\frac{1}{2}\left(s+r_{2}\right), \quad j_{\mathrm{\psi}}=\frac{1}{2}\left(r_{1}+r_{2}\right)\ .  
				\end{equation}
				Besides, the particle number can be easily get from LSH basis as
				\begin{equation}
					N_\mathrm{b}=\left(s+r_{1}\right), \quad N_\mathrm{a}=\left(s+r_{2}\right), \quad
					N_{\mathrm{\psi}}=\left(2v+r_{1}+r_{2}\right)\ .  
				\end{equation}
				
				We first calculate the action of pure boson operator $\mathcal{L}^{++}(a, b)$ as
				\begin{equation}
					\begin{aligned}
						&\mathcal{L}^{++}(a, b)\left|j_{\mathrm{b}}, j_{\mathrm{a}}, j_{\mathrm{\psi}}\right\rangle \nonumber \\
						=& \frac{1}{\sqrt{C_n}} \frac{1}{\sqrt{2 j_{\mathrm{b}}+2}}\left(\boldsymbol{a}^{\dagger} \cdot \tilde{\boldsymbol{b}}^{\dagger}\right) \frac{1}{\sqrt{2 j_{\mathrm{a}}+1}} \nonumber\\
						&\times\left(\boldsymbol{b}^{\dagger} \cdot \tilde{\boldsymbol{a}}^{\dagger}\right)^{s}\left(\boldsymbol{b}^{\dagger} \cdot \tilde{\boldsymbol{\psi}}^{\dagger}\right)^{r_{1}}\left(\boldsymbol{a}^{\dagger} \cdot \tilde{\boldsymbol{\psi}}^{\dagger}\right)^{r_{2}}\left(\boldsymbol{\psi}^{\dagger} \cdot \tilde{\boldsymbol{\psi}}^{\dagger}\right)^{v}|0\rangle
						\\
						=&-\sqrt{\frac{\left(j_{b}+j_{a}-j_{\mathrm{\psi}}+1\right)\left(j_{b}+j_{a}+j_{\mathrm{\psi}}+2\right)}{\left(2 j_{b}+2\right)\left(2 j_{a}+1\right)}}\left|j_{\mathrm{b}}+\frac{1}{2}, j_{\mathrm{a}}+\frac{1}{2}, j_{\mathrm{\psi}}\right\rangle \ ,
					\end{aligned}
				\end{equation}
				where we have used the relation $\hat{N_a}|j_{\mathrm{b}}, j_{\mathrm{a}}, j_{\mathrm{\psi}}\rangle=2j_a|j_{\mathrm{b}}, j_{\mathrm{a}}, j_{\mathrm{\psi}}\rangle,\hat{N_b}|j_{\mathrm{b}}, j_{\mathrm{a}}, j_{\mathrm{\psi}}\rangle=2j_b|j_{\mathrm{b}}, j_{\mathrm{a}}, j_{\mathrm{\psi}}\rangle$.
				Then, for operator $\mathcal{L}^{+-}(a, b)$, we have
				\begin{equation}
					\begin{aligned}
						&\mathcal{L}^{+-}(a, b)\left|j_{\mathrm{b}}, j_{\mathrm{a}}, j_{\mathrm{\psi}}\right\rangle \nonumber \\
						=&\frac{1}{\sqrt{C_n}} \frac{1}{\sqrt{2j_b}}\left(\boldsymbol{a}^{\dagger} \cdot \boldsymbol{b}\right) \frac{1}{\sqrt{2 j_{\mathrm{a}}+1}}\\
						&\times\left(\boldsymbol{b}^{\dagger} \cdot \tilde{\boldsymbol{a}}^{\dagger}\right)^{s}\left(\boldsymbol{b}^{\dagger} \cdot \tilde{\boldsymbol{\psi}}^{\dagger}\right)^{r_{1}}\left(\boldsymbol{a}^{\dagger} \cdot \tilde{\boldsymbol{\psi}}^{\dagger}\right)^{r_{2}}\left(\boldsymbol{\psi}^{\dagger} \cdot \tilde{\boldsymbol{\psi}}^{\dagger}\right)^{v}|0\rangle
						\nonumber\\
						=&\begin{cases}
							\frac{1}{\sqrt{(2 j_{\mathrm{a}}+1)2j_\mathrm{b}}}|j_{\mathrm{b}}+\frac{1}{2}, j_{\mathrm{a}}-\frac{1}{2}, j_{\mathrm{\psi}}\rangle,&r_1=1 \\
							\\
							0,&r_1=0
						\end{cases}
					\end{aligned}
				\end{equation}
				where we have used the commutation relations $\left[\left(\boldsymbol{b}^{\dagger} \cdot \boldsymbol{a}\right),\left(\boldsymbol{b}^{\dagger} \cdot \tilde{\boldsymbol{a}}^{\dagger}\right) \right]=0$ and $\left[\left(\boldsymbol{b}^{\dagger} \cdot \boldsymbol{a}\right),\left(\boldsymbol{a}^{\dagger} \cdot \tilde{\boldsymbol{\psi}}^{\dagger}\right) \right]=\left(\boldsymbol{b}^{\dagger} \cdot \tilde{\boldsymbol{\psi}}^{\dagger}\right)$. Similarly, for operator $\mathcal{L}^{-+}(a, b)$, we have
				\begin{equation}
					\begin{aligned}
						&\mathcal{L}^{-+}(a, b)\left|j_{\mathrm{b}}, j_{\mathrm{a}}, j_{\mathrm{\psi}}\right\rangle \nonumber \\
						=&-\frac{1}{\sqrt{C_n}} \frac{1}{\sqrt{2j_\mathrm{b}+2}}\left(\boldsymbol{a} \cdot \boldsymbol{b}^{\dagger}\right) \frac{1}{\sqrt{2 j_{\mathrm{a}}+1}}\\
						&\times\left(\boldsymbol{b}^{\dagger} \cdot \tilde{\boldsymbol{a}}^{\dagger}\right)^{s}\left(\boldsymbol{b}^{\dagger} \cdot \tilde{\boldsymbol{\psi}}^{\dagger}\right)^{r_{1}}\left(\boldsymbol{a}^{\dagger} \cdot \tilde{\boldsymbol{\psi}}^{\dagger}\right)^{r_{2}}\left(\boldsymbol{\psi}^{\dagger} \cdot \tilde{\boldsymbol{\psi}}^{\dagger}\right)^{v}|0\rangle
						\nonumber\\
						=&\begin{cases}
							-\frac{1}{\sqrt{(2 j_{\mathrm{b}}+2)(2j_\mathrm{a}+1)}}|j_{\mathrm{b}}-\frac{1}{2}, j_{\mathrm{a}}+\frac{1}{2}, j_{\mathrm{\psi}}\rangle,&r_2=1 \\
							\\
							0,&r_2=0
						\end{cases}
					\end{aligned}
				\end{equation}
				Finally, for operator $\mathcal{L}^{--}(a, b)$, we have
				\begin{equation}
					\begin{aligned}
						&\mathcal{L}^{--}(a, b)\left|j_{\mathrm{b}}, j_{\mathrm{a}}, j_{\mathrm{\psi}}\right\rangle \nonumber \\
						= & \frac{1}{\sqrt{C_n}} \frac{1}{\sqrt{2 j_\mathrm{b}}}\left(\boldsymbol{a} \cdot \tilde{\boldsymbol{b}}\right) \frac{1}{\sqrt{2 j_{\mathrm{a}}+1}} \\
						&\times\left(\boldsymbol{b}^{\dagger} \cdot \tilde{\boldsymbol{a}}^{\dagger}\right)^{s}\left(\boldsymbol{b}^{\dagger} \cdot \tilde{\boldsymbol{\psi}}^{\dagger}\right)^{r_{1}}\left(\boldsymbol{a}^{\dagger} \cdot \tilde{\boldsymbol{\psi}}^{\dagger}\right)^{r_{2}}\left(\boldsymbol{\psi}^{\dagger} \cdot \tilde{\boldsymbol{\psi}}^{\dagger}\right)^{v}|0\rangle\nonumber\\
						=&-\sqrt{\frac{\left(j_{\mathrm{b}}+j_{\mathrm{a}}-j_{\mathrm{\psi}}\right)\left(j_{\mathrm{b}}+j_{\mathrm{a}}+j_{\mathrm{\psi}}+1\right)}{\left(2 j_{\mathrm{a}}+1\right)\left(2 j_{\mathrm{b}}\right)}}\left|j_{\mathrm{b}}-\frac{1}{2}, j_{\mathrm{a}}-\frac{1}{2}, j_{\mathrm{\psi}}\right\rangle \ ,
					\end{aligned}
				\end{equation}
				where we have used the formulas $\left[\left(\boldsymbol{b} \cdot \tilde{\boldsymbol{a}}\right),\left(\boldsymbol{b}^{\dagger} \cdot \tilde{\boldsymbol{a}}^{\dagger}\right)  \right]=2+\hat{N_a}+\hat{N_b} $ and $s = j_a+j_b-j_\psi$.
				Note that, considering the constraint of $j_\psi$, these results are the same with that in Ref.~\cite{Hayata:2020xxm}.  
				
				Now, let us consider the actions of operators with fermion. For operator $\mathcal{S}^{++}(\psi, b)$, we have    
				\begin{equation}
					\begin{aligned}
						&\mathcal{S}^{++}(\psi, b)\left|j_{\mathrm{b}}, j_{\mathrm{a}}, j_{\mathrm{\psi}}\right\rangle \nonumber \\
						=&\frac{1}{\sqrt{C_n}} \frac{1}{\sqrt{2 j_{\mathrm{b}}+2}}\left(\boldsymbol{\psi}^{\dagger} \cdot \tilde{\boldsymbol{b}}^{\dagger}\right) \\ 
						&\times\left(\boldsymbol{b}^{\dagger} \cdot \tilde{\boldsymbol{a}}^{\dagger}\right)^{s}\left(\boldsymbol{b}^{\dagger} \cdot \tilde{\boldsymbol{\psi}}^{\dagger}\right)^{r_{1}}\left(\boldsymbol{a}^{\dagger} \cdot \tilde{\boldsymbol{\psi}}^{\dagger}\right)^{r_{2}}\left(\boldsymbol{\psi}^{\dagger} \cdot \tilde{\boldsymbol{\psi}}^{\dagger}\right)^{v}|0\rangle\nonumber\\
						=&\begin{cases}
							-\sqrt{\frac{2 j_{\mathrm{b}}+1}{2 j_{\mathrm{b}}+2}}\left|j_{\mathrm{b}}+\frac{1}{2}, j_{\mathrm{a}}, j_{\mathrm{\psi}}-\frac{1}{2}\right\rangle\nonumber,&r_2=1\\
							\\
							-\left|j_{\mathrm{b}}+\frac{1}{2}, j_{\mathrm{a}}, j_{\mathrm{\psi}}+\frac{1}{2}\right\rangle\nonumber,&r_1=r_2=v=0\\
							\\
							0, &\text{other cases}
						\end{cases}
					\end{aligned}
				\end{equation}
				where we have used the Eq.~\eqref{usefuleq}. Similarly, for operator $\mathcal{S}^{++}(\psi, a)$, we have
				\begin{equation}
					\begin{aligned}
						&\mathcal{S}^{++}(\psi, a)\left|j_{\mathrm{b}}, j_{\mathrm{a}}, j_{\mathrm{\psi}}\right\rangle \nonumber \\
						=&\frac{1}{\sqrt{C_n}} \frac{1}{\sqrt{2 j_{\mathrm{a}}+2}}\left(\boldsymbol{\psi}^{\dagger} \cdot \tilde{\boldsymbol{a}}^{\dagger}\right) \\ 
						&\times\left(\boldsymbol{b}^{\dagger} \cdot \tilde{\boldsymbol{a}}^{\dagger}\right)^{s}\left(\boldsymbol{b}^{\dagger} \cdot \tilde{\boldsymbol{\psi}}^{\dagger}\right)^{r_{1}}\left(\boldsymbol{a}^{\dagger} \cdot \tilde{\boldsymbol{\psi}}^{\dagger}\right)^{r_{2}}\left(\boldsymbol{\psi}^{\dagger} \cdot \tilde{\boldsymbol{\psi}}^{\dagger}\right)^{v}|0\rangle\nonumber \\
						=&\begin{cases}
							\sqrt{\frac{2 j_{\mathrm{a}}+1}{2 j_{\mathrm{a}}+2}}\left|j_{\mathrm{b}}, j_{\mathrm{a}}+\frac{1}{2}, j_{\mathrm{\psi}}-\frac{1}{2}\right\rangle\nonumber ,&r_1=1\\
							\\
							-\left|j_{\mathrm{b}}, j_{\mathrm{a}}+\frac{1}{2}, j_{\mathrm{\psi}}+\frac{1}{2}\right\rangle\nonumber ,&r_1=r_2=v=0\\
							\\
							0,&\text{other cases}
						\end{cases}
					\end{aligned}
				\end{equation}
				Then, for operator $\mathcal{S}^{+-}(b, \psi)$, we have
				\begin{equation}
					\begin{aligned}
						&\mathcal{S}^{+-}(b, \psi)\left|j_{\mathrm{b}}, j_{\mathrm{a}}, j_{\mathrm{\psi}}\right\rangle \nonumber \\
						=&\frac{1}{\sqrt{C_n}} \frac{1}{\sqrt{\left(2 j_{\mathrm{b}}+1\right)}}\left(\boldsymbol{b}^{\dagger} \cdot \boldsymbol{\psi}\right)\\ 
						& \times\left(\boldsymbol{b}^{\dagger} \cdot \tilde{\boldsymbol{a}}^{\dagger}\right)^{s}\left(\boldsymbol{b}^{\dagger} \cdot \tilde{\boldsymbol{\psi}}^{\dagger}\right)^{r_{1}}\left(\boldsymbol{a}^{\dagger} \cdot \tilde{\boldsymbol{\psi}}^{\dagger}\right)^{r_{2}}\left(\boldsymbol{\psi}^{\dagger} \cdot \tilde{\boldsymbol{\psi}}^{\dagger}\right)^{v}|0\rangle\nonumber \\
						=&\begin{cases}
							-\left|j_{\mathrm{b}}+\frac{1}{2}, j_{\mathrm{a}}, j_{\mathrm{\psi}}-\frac{1}{2}\right\rangle\nonumber,&r_2=1\\
							\\
							\sqrt{\frac{2 j_{\mathrm{b}}+2}{2 j_{\mathrm{b}}+1}}\left|j_{\mathrm{b}}+\frac{1}{2}, j_{\mathrm{a}}, j_{\mathrm{\psi}}+\frac{1}{2}\right\rangle\nonumber ,&v=1\\
							\\
							0,& \text{other cases}
						\end{cases}
					\end{aligned}
				\end{equation}
				where we have used the commutation relation $\left[ \left(\boldsymbol{b}^{\dagger} \cdot \boldsymbol{\psi}\right),\left(\boldsymbol{a}^{\dagger} \cdot \tilde{\boldsymbol{\psi}}^{\dagger}\right)\right]=- \left(\boldsymbol{b}^{\dagger} \cdot \tilde{\boldsymbol{a}}^{\dagger}\right)$.
				Similarly, for operator $\mathcal{S}^{+-}(a, \psi)$, we have
				\begin{equation}
					\begin{aligned}
						&\mathcal{S}^{+-}(a, \psi)\left|j_{\mathrm{b}}, j_{\mathrm{a}}, j_{\mathrm{\psi}}\right\rangle \nonumber \\
						=&\frac{1}{\sqrt{C_n}} \frac{1}{\sqrt{\left(2 j_{\mathrm{b}}+1\right)}}\left(\boldsymbol{a}^{\dagger} \cdot \boldsymbol{\psi}\right)\\ 
						\times& \left(\boldsymbol{b}^{\dagger} \cdot \tilde{\boldsymbol{a}}^{\dagger}\right)^{s}\left(\boldsymbol{b}^{\dagger} \cdot \tilde{\boldsymbol{\psi}}^{\dagger}\right)^{r_{1}}\left(\boldsymbol{a}^{\dagger} \cdot \tilde{\boldsymbol{\psi}}^{\dagger}\right)^{r_{2}}\left(\boldsymbol{\psi}^{\dagger} \cdot \tilde{\boldsymbol{\psi}}^{\dagger}\right)^{v}|0\rangle
						\nonumber \\
						=&\begin{cases}
							\left|j_{\mathrm{b}}, j_{\mathrm{a}}+\frac{1}{2}, j_{\mathrm{\psi}}-\frac{1}{2}\right\rangle\nonumber ,&r_1=1\\
							\\
							\sqrt{\frac{2 j_{\mathrm{a}}+2}{2 j_{\mathrm{a}}+1}}\left|j_{\mathrm{b}}, j_{\mathrm{a}}+\frac{1}{2}, j_{\mathrm{\psi}}+\frac{1}{2}\right\rangle\nonumber,&v=1 \\
							\\
							0, &\text{other cases}
						\end{cases}
					\end{aligned}
				\end{equation}
				These actions of gauge invariant operators are consistent with the results of the Ref.~\cite{anishetty2019gauss}.

				\section{Hamiltonian matrix under spin basis}\label{B}
				\subsection{Exact mapping to spin system}\label{B1}
				Here, we list all the correspondences between the basis states of the LSH representation and the physical states encoded in six qubits. In Sec.~\ref{Sec3}, we use the first three qubits to represent groups which the physical states belong to, namely
				\begin{equation}
					\begin{aligned}
						\mathrm{group~A}:\qquad& |\uparrow\rangle|\uparrow\rangle|\uparrow\rangle \ ,\\
						\mathrm{group~B}:\qquad& |\uparrow\rangle|\uparrow\rangle|\downarrow\rangle \ ,\\
						\mathrm{group~C_{1}}:\qquad& |\uparrow\rangle|\downarrow\rangle|\uparrow\rangle \ ,\\
						\mathrm{group~C_{2}}:\qquad & |\uparrow\rangle|\downarrow\rangle|\downarrow\rangle \ ,\\
						\mathrm{group~D}:  \qquad   & |\downarrow\rangle|\uparrow\rangle|\uparrow\rangle \ ,\\
						\mathrm{group~E}:  \qquad   & |\downarrow\rangle|\uparrow\rangle|\downarrow\rangle \ .
					\end{aligned}
				\end{equation}
				where the groups A, B, C, D, and E correspond to the
				global states with 0, 1, 2, 3, and 4 hats, respectively.
				For all physical states mentioned in Sec.~\ref{22}, we show the exact mapping from the physical states to spin basis with three other qubits.
				For the group A, we have
				\begin{equation}
					\begin{aligned}
						&|0220\rangle = |\uparrow\rangle|\uparrow\rangle|\uparrow\rangle,\\
						&|2200\rangle = |\uparrow\rangle|\uparrow\rangle|\downarrow\rangle,\\
						&|2002\rangle = |\uparrow\rangle|\downarrow\rangle|\uparrow\rangle,\\
						&|0022\rangle = |\uparrow\rangle|\downarrow\rangle|\downarrow\rangle,\\
						&|2020\rangle = |\downarrow\rangle|\uparrow\rangle|\uparrow\rangle,\\
						&|0202\rangle = |\downarrow\rangle|\uparrow\rangle|\downarrow\rangle, 
					\end{aligned}
				\end{equation}
				where we omit to write the first three qubits, $|\uparrow\rangle|\uparrow\rangle|\uparrow\rangle$ which labels the group A. Later, we always omit the first three qubits which label the group. 
				For the group B, we have
				\begin{equation}
					\begin{aligned}
						&|\hat{1}120\rangle = |\uparrow\rangle|\uparrow\rangle|\uparrow\rangle,\\
						&|120\hat{1}\rangle = |\uparrow\rangle|\uparrow\rangle|\downarrow\rangle,\\
						&|20\hat{1}1\rangle = |\uparrow\rangle|\downarrow\rangle|\uparrow\rangle,\\
						&|0\hat{1}12\rangle = |\uparrow\rangle|\downarrow\rangle|\downarrow\rangle,\\
						&|02\hat{1}1\rangle = |\downarrow\rangle|\uparrow\rangle|\uparrow\rangle,\\
						&|2\hat{1}10\rangle = |\downarrow\rangle|\uparrow\rangle|\downarrow\rangle,\\
						&|\hat{1}102\rangle = |\downarrow\rangle|\downarrow\rangle|\uparrow\rangle,\\
						&|102\hat{1}\rangle = |\downarrow\rangle|\downarrow\rangle|\downarrow\rangle.
					\end{aligned}
				\end{equation}
				For the group $\mathrm{C_1}$, we have
				\begin{equation}
					\begin{aligned}
						&|\hat{1}\hat{2}10\rangle = |\uparrow\rangle|\uparrow\rangle|\uparrow\rangle,\\
						&|\hat{2}10\hat{1}\rangle = |\uparrow\rangle|\uparrow\rangle|\downarrow\rangle,\\
						&|10\hat{1}\hat{2}\rangle = |\uparrow\rangle|\downarrow\rangle|\uparrow\rangle,\\
						&|0\hat{1}\hat{2}1\rangle = |\uparrow\rangle|\downarrow\rangle|\downarrow\rangle,\\
						&|\hat{0}12\hat{1}\rangle = |\downarrow\rangle|\uparrow\rangle|\uparrow\rangle,\\
						&|12\hat{1}\hat{0}\rangle = |\downarrow\rangle|\uparrow\rangle|\downarrow\rangle,\\
						&|2\hat{1}\hat{0}1\rangle = |\downarrow\rangle|\downarrow\rangle|\uparrow\rangle,\\
						&|\hat{1}\hat{0}12\rangle = |\downarrow\rangle|\downarrow\rangle|\downarrow\rangle.
					\end{aligned}
				\end{equation}
				For the group $\mathrm{C_2}$, we have
				\begin{equation}
					\begin{aligned}
						&|\hat{1}1\hat{1}1\rangle = |\uparrow\rangle|\uparrow\rangle|\uparrow\rangle,\\
						&|1\hat{1}1\hat{1}\rangle = |\uparrow\rangle|\uparrow\rangle|\downarrow\rangle.
					\end{aligned}
				\end{equation}
				For the group D, we have
				\begin{equation}
					\begin{aligned}
						&|\hat{1}\hat{2}\hat{0}1\rangle = |\uparrow\rangle|\uparrow\rangle|\uparrow\rangle,\\
						&|\hat{2}\hat{0}1\hat{1}\rangle = |\uparrow\rangle|\uparrow\rangle|\downarrow\rangle,\\
						&|\hat{0}1\hat{1}\hat{2}\rangle = |\uparrow\rangle|\downarrow\rangle|\uparrow\rangle,\\
						&|1\hat{1}\hat{2}\hat{0}\rangle = |\uparrow\rangle|\downarrow\rangle|\downarrow\rangle,\\
						&|\hat{0}\hat{2}1\hat{1}\rangle = |\downarrow\rangle|\uparrow\rangle|\uparrow\rangle,\\
						&|\hat{2}1\hat{1}\hat{0}\rangle = |\downarrow\rangle|\uparrow\rangle|\downarrow\rangle,\\
						&|1\hat{1}\hat{0}\hat{2}\rangle = |\downarrow\rangle|\downarrow\rangle|\uparrow\rangle,\\
						&|\hat{1}\hat{0}\hat{2}1\rangle = |\downarrow\rangle|\downarrow\rangle|\downarrow\rangle.
					\end{aligned}
				\end{equation}
				For the group E, we have
				\begin{equation}
					\begin{aligned}
						&|\hat{2}\hat{2}\hat{0}\hat{0}\rangle = |\uparrow\rangle|\uparrow\rangle|\uparrow\rangle,\\
						&|\hat{2}\hat{0}\hat{0}\hat{2}\rangle = |\uparrow\rangle|\uparrow\rangle|\downarrow\rangle,\\
						&|\hat{0}\hat{0}\hat{2}\hat{2}\rangle = |\uparrow\rangle|\downarrow\rangle|\uparrow\rangle,\\
						&|\hat{0}\hat{2}\hat{2}\hat{0}\rangle = |\uparrow\rangle|\downarrow\rangle|\downarrow\rangle,\\
						&|\hat{2}\hat{0}\hat{2}\hat{0}\rangle = |\downarrow\rangle|\uparrow\rangle|\uparrow\rangle,\\
						&|\hat{0}\hat{2}\hat{0}\hat{2}\rangle = |\downarrow\rangle|\uparrow\rangle|\downarrow\rangle.
					\end{aligned}
				\end{equation}

				\subsection{All terms of Hamiltonian matrix}\label{B2}
				In Sec.~\ref{Sec3}, we have showed the approach to get the Hamiltonian matrix through an example. Here, according to the actions shown in Sec.~\ref{23}, we give all terms of the Hamiltonian matrix using Pauli matrices.
				For electric part, we have
				\begin{equation}
					\begin{aligned}
						H_{\rm E}& =\frac{3}{32}(I+Z)(I+Z)(I-Z)III\\
						& +\frac{3}{16}(I+Z)(I-Z)(I+Z)III\\
						&+\frac{3}{64}(I+Z)(I-Z)(I-Z)(I+Z)(I+Z)I\\
						&+\frac{9}{32}(I-Z)(I+Z)(I+Z)III\\
						&+\frac{3}{16}(I-Z)(I+Z)(I-Z)(I+Z)II\\
						&+\frac{3}{32}(I-Z)(I+Z)(I-Z)(I-Z)(I+Z)I,
					\end{aligned}   
				\end{equation}
				where we omit the sign $\otimes$ for simplicity. Later, we always do this.
				For magnetic part, we have
				\begin{equation}
					\begin{aligned}
						H_{\rm B}&= \frac{1}{8}[X(I+Z)X-Y(I+Z)Y](I-Z)(I+Z)X\\
						&-\frac{1}{8}[X(I+Z)X+Y(I+Z)Y](I-Z)II\\
						&-\frac{1}{8}Y(I+Z)Y(I+Z)XX
						-\frac{1}{8}X(I+Z)Y(I+Z)XY\\
						&-\frac{1}{8}Y(I+Z)Y(I+Z)IX
						+\frac{1}{8}X(I+Z)Y(I+Z)IY\\
						&+\frac{1}{16}[X(I+Z)X-Y(I+Z)Y](I+Z)(XX+IX)\\
						&+\frac{1}{16}[Y(I+Z)X+X(I+Z)Y](I+Z)(IY-XY)\\
						&-\frac{1}{16}(I+Z)(I-Z)(I+Z)(XIX-YIY)\\
						&-\frac{1}{16}(I+Z)(I-Z)(I+Z)(XXX+YXY)\\
						&+\frac{1}{64}(I+Z)(I-Z)(I-Z)(I+Z)(I+Z)X.
					\end{aligned}
				\end{equation}
				For fermion mass part, we have
				\begin{equation}
					\begin{aligned}
						H_{\rm FM} &=\frac{m}{8}(I-Z)(I+Z)(I-Z)(I-Z)(I+Z)Z\\
						&-\frac{m}{8}(I+Z)(I+Z)(I+Z)(I-Z)(I+Z)Z\\
						&-\frac{m}{4}(I+Z)(I+Z)(I-Z)ZIZ\\
						&+\frac{m}{4}(I-Z)(I+Z)(I+Z)IIZ.
					\end{aligned}
				\end{equation}
				Finally, for fermion kinetic part, we first divide it into different parts as
				\begin{equation}
					H_{\rm FK}=H_{\rm AB}+H_{\rm BC}+H_{\rm CD}+H_{\rm DE}
				\end{equation}
				where $ H_{\rm AB}$ represents the Hamiltonian matrix based on the actions which transport the group A to group B and the others are similar. 
				For $ H_{\rm AB}$, we have
				\begin{equation}
					\begin{aligned}
						H_{\rm AB}=&-\frac{\sqrt{2}}{16}(I+Z)(I+Z)X(I-Z)(I+Z+X)X\\
						&+\frac{\sqrt{2}}{16}(I+Z)(I+Z)Y(I-Z)YX\\
						&-\frac{\sqrt{2}}{16}(I+Z)(I+Z)(XX+YY)(I+Z+X)I\\
						&-\frac{\sqrt{2}}{16}(I+Z)(XY-YX)YI\\
						&-\frac{\sqrt{2}}{8}(I+Z)(I+Z)(XI+XZ+XX-YY)II.\\
					\end{aligned}
				\end{equation}
				For $ H_{\rm BC}$, we have 
				\begin{equation}
					\begin{aligned}
						H_{\rm BC}&=\frac{1}{8}
						(I+Z)(XX+YY)(I+Z-X)II\\
						&+\frac{1}{8}
						(I+Z)(YX-XY)YII\\
						&+\frac{1}{16}
						(I+Z)(XX+YY)(XXX+YXY)\\
						&+\frac{1}{16}(I+Z)(YX-XY)(YXX-XXY)\\
						&+\frac{1}{16}
						(I+Z)(XX+YY)(XIX-YIY)\\
						&+\frac{1}{16}(I+Z)(YX-XY)(YIX+XIY)\\
						&-\frac{1}{16}
						(I+Z)(XX+YY)(I-Z)IX\\
						&+\frac{1}{16}(I+Z)(XY-YX)(I-Z)IY\\
						&-\frac{1}{16}
						(I+Z)(XX+YY)(I-Z)XX\\
						&-\frac{1}{16}(I+Z)(XY-YX)(I-Z)XY\\
						&-\frac{\sqrt{2}}{16}
						(I+Z)X(I-Z)(I+Z+X)(I+Z+X)I\\
						&-\frac{\sqrt{2}}{16}(I+Z)X(I-Z)YYI\\
						&+\frac{\sqrt{2}}{16}(I+Z)Y(I-Z)(I+Z+X)YI\\
						&+\frac{\sqrt{2}}{16}(I+Z)Y(I-Z)Y(I+Z+X)I.
					\end{aligned}
				\end{equation}
				For $ H_{\rm CD}$, we have
				\begin{equation}
					\begin{aligned}
						H_{\rm CD}&=\frac{1}{8}
						(XX+YY)(I+Z)(I-Z)II\\
						&-\frac{1}{8}
						(XX+YY)(I+Z)(I+Z+X)II\\
						&-\frac{1}{8}(XY-YX)(I+Z)YII\\
						&+\frac{1}{8}
						(XX+YY)(I+Z)XXI\\
						&+\frac{1}{8}(YX-XY)(I+Z)Y)XI\\
						&+\frac{\sqrt{2}}{32}(XX+YY)(XX+YY)(I+Z+X)X\\
						&+\frac{\sqrt{2}}{32}(YX-XY)(YX-XY)(I+Z+X)X\\
						&+\frac{\sqrt{2}}{32}(XX+YY)(YX-XY)YX\\
						&-\frac{\sqrt{2}}{32}(YX-XY)(XX+YY)YX\\
						&+\frac{\sqrt{2}}{32}(XX+YY)X(I+Z)(I+Z+X)I\\
						&+\frac{\sqrt{2}}{32}(XX+YY)Y(I+Z)YI\\
						&+\frac{\sqrt{2}}{32}(YX-XY)Y(I+Z)(I+Z+X)I\\
						&-\frac{\sqrt{2}}{32}(YX-XY)X(I+Z)YI.
					\end{aligned}
				\end{equation}
				For $ H_{\rm DE}$, we have
				\begin{equation}
					\begin{aligned}
						H_{\rm DE}&=\frac{\sqrt{2}}{16}
						(I-Z)(I+Z)X(I+Z)II\\
						&+\frac{\sqrt{2}}{32}
						(I-Z)(I+Z)X(I-Z+X)(X+I+Z)I\\
						&-\frac{\sqrt{2}}{32}(I-Z)(I+Z)YY(X+I+Z)I\\
						&+\frac{\sqrt{2}}{32}
						(I-Z)(I+Z)XYYI\\
						&+\frac{\sqrt{2}}{32}(I-Z)(I+Z)Y(I+X-Z)YI\\
						&+\frac{\sqrt{2}}{32}
						(I-Z)(I+Z)(XX+YY)XX\\
						&+\frac{\sqrt{2}}{32}(I-Z)(I+Z)(XY-YX)XY\\
						&+\frac{\sqrt{2}}{32}
						(I-Z)(I+Z)(XX+YY)IX\\
						&-\frac{\sqrt{2}}{32}(I-Z)(I+Z)(XY-YX)IY.
					\end{aligned}
				\end{equation}
				\clearpage
				\bibliography{main}

\begin{thebibliography}{66}%
\makeatletter
\providecommand \@ifxundefined [1]{%
 \@ifx{#1\undefined}
}%
\providecommand \@ifnum [1]{%
 \ifnum #1\expandafter \@firstoftwo
 \else \expandafter \@secondoftwo
 \fi
}%
\providecommand \@ifx [1]{%
 \ifx #1\expandafter \@firstoftwo
 \else \expandafter \@secondoftwo
 \fi
}%
\providecommand \natexlab [1]{#1}%
\providecommand \enquote  [1]{``#1''}%
\providecommand \bibnamefont  [1]{#1}%
\providecommand \bibfnamefont [1]{#1}%
\providecommand \citenamefont [1]{#1}%
\providecommand \href@noop [0]{\@secondoftwo}%
\providecommand \href [0]{\begingroup \@sanitize@url \@href}%
\providecommand \@href[1]{\@@startlink{#1}\@@href}%
\providecommand \@@href[1]{\endgroup#1\@@endlink}%
\providecommand \@sanitize@url [0]{\catcode `\\12\catcode `\$12\catcode `\&12\catcode `\#12\catcode `\^12\catcode `\_12\catcode `\%12\relax}%
\providecommand \@@startlink[1]{}%
\providecommand \@@endlink[0]{}%
\providecommand \url  [0]{\begingroup\@sanitize@url \@url }%
\providecommand \@url [1]{\endgroup\@href {#1}{\urlprefix }}%
\providecommand \urlprefix  [0]{URL }%
\providecommand \Eprint [0]{\href }%
\providecommand \doibase [0]{https://doi.org/}%
\providecommand \selectlanguage [0]{\@gobble}%
\providecommand \bibinfo  [0]{\@secondoftwo}%
\providecommand \bibfield  [0]{\@secondoftwo}%
\providecommand \translation [1]{[#1]}%
\providecommand \BibitemOpen [0]{}%
\providecommand \bibitemStop [0]{}%
\providecommand \bibitemNoStop [0]{.\EOS\space}%
\providecommand \EOS [0]{\spacefactor3000\relax}%
\providecommand \BibitemShut  [1]{\csname bibitem#1\endcsname}%
\let\auto@bib@innerbib\@empty
\bibitem [{\citenamefont {Wilson}(1974)}]{Wilson:1974sk}%
  \BibitemOpen
  \bibfield  {author} {\bibinfo {author} {\bibfnamefont {K.~G.}\ \bibnamefont {Wilson}},\ }\bibfield  {title} {\bibinfo {title} {{Confinement of Quarks}},\ }\href {https://doi.org/10.1103/PhysRevD.10.2445} {\bibfield  {journal} {\bibinfo  {journal} {Phys. Rev. D}\ }\textbf {\bibinfo {volume} {10}},\ \bibinfo {pages} {2445} (\bibinfo {year} {1974})}\BibitemShut {NoStop}%
\bibitem [{\citenamefont {Troyer}\ and\ \citenamefont {Wiese}(2005)}]{PhysRevLett.94.170201}%
  \BibitemOpen
  \bibfield  {author} {\bibinfo {author} {\bibfnamefont {M.}~\bibnamefont {Troyer}}\ and\ \bibinfo {author} {\bibfnamefont {U.-J.}\ \bibnamefont {Wiese}},\ }\bibfield  {title} {\bibinfo {title} {Computational complexity and fundamental limitations to fermionic quantum monte carlo simulations},\ }\href {https://doi.org/10.1103/PhysRevLett.94.170201} {\bibfield  {journal} {\bibinfo  {journal} {Phys. Rev. Lett.}\ }\textbf {\bibinfo {volume} {94}},\ \bibinfo {pages} {170201} (\bibinfo {year} {2005})}\BibitemShut {NoStop}%
\bibitem [{\citenamefont {de~Forcrand}(2009)}]{deForcrand:2009zkb}%
  \BibitemOpen
  \bibfield  {author} {\bibinfo {author} {\bibfnamefont {P.}~\bibnamefont {de~Forcrand}},\ }\bibfield  {title} {\bibinfo {title} {{Simulating QCD at finite density}},\ }\href {https://doi.org/10.22323/1.091.0010} {\bibfield  {journal} {\bibinfo  {journal} {PoS}\ }\textbf {\bibinfo {volume} {LAT2009}},\ \bibinfo {pages} {010} (\bibinfo {year} {2009})},\ \Eprint {https://arxiv.org/abs/1005.0539} {arXiv:1005.0539 [hep-lat]} \BibitemShut {NoStop}%
\bibitem [{\citenamefont {Gattringer}\ and\ \citenamefont {Langfeld}(2016)}]{Gattringer:2016kco}%
  \BibitemOpen
  \bibfield  {author} {\bibinfo {author} {\bibfnamefont {C.}~\bibnamefont {Gattringer}}\ and\ \bibinfo {author} {\bibfnamefont {K.}~\bibnamefont {Langfeld}},\ }\bibfield  {title} {\bibinfo {title} {{Approaches to the sign problem in lattice field theory}},\ }\href {https://doi.org/10.1142/S0217751X16430077} {\bibfield  {journal} {\bibinfo  {journal} {Int. J. Mod. Phys. A}\ }\textbf {\bibinfo {volume} {31}},\ \bibinfo {pages} {1643007} (\bibinfo {year} {2016})},\ \Eprint {https://arxiv.org/abs/1603.09517} {arXiv:1603.09517 [hep-lat]} \BibitemShut {NoStop}%
\bibitem [{\citenamefont {Nagata}(2020)}]{Nagata:2021bru}%
  \BibitemOpen
  \bibfield  {author} {\bibinfo {author} {\bibfnamefont {K.}~\bibnamefont {Nagata}},\ }\bibfield  {title} {\bibinfo {title} {{Finite-density lattice QCD and sign problem: current status and open problems}},\ }\href {https://www2.yukawa.kyoto-u.ac.jp/~soken.editorial/sokendenshi/vol31/Nagata2016GKH_EI_revised.pdf} {\bibfield  {journal} {\bibinfo  {journal} {Soryusironkenkyu}\ }\textbf {\bibinfo {volume} {31}},\ \bibinfo {pages} {1} (\bibinfo {year} {2020})}\BibitemShut {NoStop}%
\bibitem [{\citenamefont {Nagata}(2022)}]{Nagata:2021ugx}%
  \BibitemOpen
  \bibfield  {author} {\bibinfo {author} {\bibfnamefont {K.}~\bibnamefont {Nagata}},\ }\bibfield  {title} {\bibinfo {title} {{Finite-density lattice QCD and sign problem: Current status and open problems}},\ }\href {https://doi.org/10.1016/j.ppnp.2022.103991} {\bibfield  {journal} {\bibinfo  {journal} {Prog. Part. Nucl. Phys.}\ }\textbf {\bibinfo {volume} {127}},\ \bibinfo {pages} {103991} (\bibinfo {year} {2022})},\ \Eprint {https://arxiv.org/abs/2108.12423} {arXiv:2108.12423 [hep-lat]} \BibitemShut {NoStop}%
\bibitem [{\citenamefont {Parisi}\ and\ \citenamefont {Wu}(1981)}]{Parisi:1980ys}%
  \BibitemOpen
  \bibfield  {author} {\bibinfo {author} {\bibfnamefont {G.}~\bibnamefont {Parisi}}\ and\ \bibinfo {author} {\bibfnamefont {Y.-s.}\ \bibnamefont {Wu}},\ }\bibfield  {title} {\bibinfo {title} {{Perturbation Theory Without Gauge Fixing}},\ }\href@noop {} {\bibfield  {journal} {\bibinfo  {journal} {Sci.Sin.}\ }\textbf {\bibinfo {volume} {24}},\ \bibinfo {pages} {483} (\bibinfo {year} {1981})}\BibitemShut {NoStop}%
\bibitem [{\citenamefont {Klauder}(1984)}]{Klauder:1983sp}%
  \BibitemOpen
  \bibfield  {author} {\bibinfo {author} {\bibfnamefont {J.~R.}\ \bibnamefont {Klauder}},\ }\bibfield  {title} {\bibinfo {title} {{Coherent State Langevin Equations for Canonical Quantum Systems With Applications to the Quantized Hall Effect}},\ }\href {https://doi.org/10.1103/PhysRevA.29.2036} {\bibfield  {journal} {\bibinfo  {journal} {Phys. Rev. A}\ }\textbf {\bibinfo {volume} {29}},\ \bibinfo {pages} {2036} (\bibinfo {year} {1984})}\BibitemShut {NoStop}%
\bibitem [{\citenamefont {Parisi}(1983)}]{Parisi:1983mgm}%
  \BibitemOpen
  \bibfield  {author} {\bibinfo {author} {\bibfnamefont {G.}~\bibnamefont {Parisi}},\ }\bibfield  {title} {\bibinfo {title} {{ON COMPLEX PROBABILITIES}},\ }\href {https://doi.org/10.1016/0370-2693(83)90525-7} {\bibfield  {journal} {\bibinfo  {journal} {Phys. Lett. B}\ }\textbf {\bibinfo {volume} {131}},\ \bibinfo {pages} {393} (\bibinfo {year} {1983})}\BibitemShut {NoStop}%
\bibitem [{\citenamefont {Witten}(2011)}]{Witten:2010cx}%
  \BibitemOpen
  \bibfield  {author} {\bibinfo {author} {\bibfnamefont {E.}~\bibnamefont {Witten}},\ }\bibfield  {title} {\bibinfo {title} {{Analytic Continuation Of Chern-Simons Theory}},\ }\href@noop {} {\bibfield  {journal} {\bibinfo  {journal} {AMS/IP Stud. Adv. Math.}\ }\textbf {\bibinfo {volume} {50}},\ \bibinfo {pages} {347} (\bibinfo {year} {2011})},\ \Eprint {https://arxiv.org/abs/1001.2933} {arXiv:1001.2933 [hep-th]} \BibitemShut {NoStop}%
\bibitem [{\citenamefont {Cristoforetti}\ \emph {et~al.}(2012)\citenamefont {Cristoforetti}, \citenamefont {Di~Renzo},\ and\ \citenamefont {Scorzato}}]{Cristoforetti:2012su}%
  \BibitemOpen
  \bibfield  {author} {\bibinfo {author} {\bibfnamefont {M.}~\bibnamefont {Cristoforetti}}, \bibinfo {author} {\bibfnamefont {F.}~\bibnamefont {Di~Renzo}},\ and\ \bibinfo {author} {\bibfnamefont {L.}~\bibnamefont {Scorzato}} (\bibinfo {collaboration} {AuroraScience}),\ }\bibfield  {title} {\bibinfo {title} {{New approach to the sign problem in quantum field theories: High density QCD on a Lefschetz thimble}},\ }\href {https://doi.org/10.1103/PhysRevD.86.074506} {\bibfield  {journal} {\bibinfo  {journal} {Phys. Rev. D}\ }\textbf {\bibinfo {volume} {86}},\ \bibinfo {pages} {074506} (\bibinfo {year} {2012})},\ \Eprint {https://arxiv.org/abs/1205.3996} {arXiv:1205.3996 [hep-lat]} \BibitemShut {NoStop}%
\bibitem [{\citenamefont {Fujii}\ \emph {et~al.}(2013)\citenamefont {Fujii}, \citenamefont {Honda}, \citenamefont {Kato}, \citenamefont {Kikukawa}, \citenamefont {Komatsu},\ and\ \citenamefont {Sano}}]{Fujii:2013sra}%
  \BibitemOpen
  \bibfield  {author} {\bibinfo {author} {\bibfnamefont {H.}~\bibnamefont {Fujii}}, \bibinfo {author} {\bibfnamefont {D.}~\bibnamefont {Honda}}, \bibinfo {author} {\bibfnamefont {M.}~\bibnamefont {Kato}}, \bibinfo {author} {\bibfnamefont {Y.}~\bibnamefont {Kikukawa}}, \bibinfo {author} {\bibfnamefont {S.}~\bibnamefont {Komatsu}},\ and\ \bibinfo {author} {\bibfnamefont {T.}~\bibnamefont {Sano}},\ }\bibfield  {title} {\bibinfo {title} {{Hybrid Monte Carlo on Lefschetz thimbles - A study of the residual sign problem}},\ }\href {https://doi.org/10.1007/JHEP10(2013)147} {\bibfield  {journal} {\bibinfo  {journal} {JHEP}\ }\textbf {\bibinfo {volume} {10}},\ \bibinfo {pages} {147}},\ \Eprint {https://arxiv.org/abs/1309.4371} {arXiv:1309.4371 [hep-lat]} \BibitemShut {NoStop}%
\bibitem [{\citenamefont {Mori}\ \emph {et~al.}(2017)\citenamefont {Mori}, \citenamefont {Kashiwa},\ and\ \citenamefont {Ohnishi}}]{Mori:2017pne}%
  \BibitemOpen
  \bibfield  {author} {\bibinfo {author} {\bibfnamefont {Y.}~\bibnamefont {Mori}}, \bibinfo {author} {\bibfnamefont {K.}~\bibnamefont {Kashiwa}},\ and\ \bibinfo {author} {\bibfnamefont {A.}~\bibnamefont {Ohnishi}},\ }\bibfield  {title} {\bibinfo {title} {{Toward solving the sign problem with path optimization method}},\ }\href {https://doi.org/10.1103/PhysRevD.96.111501} {\bibfield  {journal} {\bibinfo  {journal} {Phys. Rev. D}\ }\textbf {\bibinfo {volume} {96}},\ \bibinfo {pages} {111501} (\bibinfo {year} {2017})},\ \Eprint {https://arxiv.org/abs/1705.05605} {arXiv:1705.05605 [hep-lat]} \BibitemShut {NoStop}%
\bibitem [{\citenamefont {Mori}\ \emph {et~al.}(2018)\citenamefont {Mori}, \citenamefont {Kashiwa},\ and\ \citenamefont {Ohnishi}}]{Mori:2017nwj}%
  \BibitemOpen
  \bibfield  {author} {\bibinfo {author} {\bibfnamefont {Y.}~\bibnamefont {Mori}}, \bibinfo {author} {\bibfnamefont {K.}~\bibnamefont {Kashiwa}},\ and\ \bibinfo {author} {\bibfnamefont {A.}~\bibnamefont {Ohnishi}},\ }\bibfield  {title} {\bibinfo {title} {{Application of a neural network to the sign problem via the path optimization method}},\ }\href {https://doi.org/10.1093/ptep/ptx191} {\bibfield  {journal} {\bibinfo  {journal} {PTEP}\ }\textbf {\bibinfo {volume} {2018}},\ \bibinfo {pages} {023B04} (\bibinfo {year} {2018})},\ \Eprint {https://arxiv.org/abs/1709.03208} {arXiv:1709.03208 [hep-lat]} \BibitemShut {NoStop}%
\bibitem [{\citenamefont {Alexandru}\ \emph {et~al.}(2018)\citenamefont {Alexandru}, \citenamefont {Bedaque}, \citenamefont {Lamm},\ and\ \citenamefont {Lawrence}}]{Alexandru:2018fqp}%
  \BibitemOpen
  \bibfield  {author} {\bibinfo {author} {\bibfnamefont {A.}~\bibnamefont {Alexandru}}, \bibinfo {author} {\bibfnamefont {P.~F.}\ \bibnamefont {Bedaque}}, \bibinfo {author} {\bibfnamefont {H.}~\bibnamefont {Lamm}},\ and\ \bibinfo {author} {\bibfnamefont {S.}~\bibnamefont {Lawrence}},\ }\bibfield  {title} {\bibinfo {title} {{Finite-Density Monte Carlo Calculations on Sign-Optimized Manifolds}},\ }\href {https://doi.org/10.1103/PhysRevD.97.094510} {\bibfield  {journal} {\bibinfo  {journal} {Phys. Rev. D}\ }\textbf {\bibinfo {volume} {97}},\ \bibinfo {pages} {094510} (\bibinfo {year} {2018})},\ \Eprint {https://arxiv.org/abs/1804.00697} {arXiv:1804.00697 [hep-lat]} \BibitemShut {NoStop}%
\bibitem [{\citenamefont {Namekawa}\ \emph {et~al.}(2023)\citenamefont {Namekawa}, \citenamefont {Kashiwa}, \citenamefont {Matsuda}, \citenamefont {Ohnishi},\ and\ \citenamefont {Takase}}]{Namekawa:2022liz}%
  \BibitemOpen
  \bibfield  {author} {\bibinfo {author} {\bibfnamefont {Y.}~\bibnamefont {Namekawa}}, \bibinfo {author} {\bibfnamefont {K.}~\bibnamefont {Kashiwa}}, \bibinfo {author} {\bibfnamefont {H.}~\bibnamefont {Matsuda}}, \bibinfo {author} {\bibfnamefont {A.}~\bibnamefont {Ohnishi}},\ and\ \bibinfo {author} {\bibfnamefont {H.}~\bibnamefont {Takase}},\ }\bibfield  {title} {\bibinfo {title} {{Improving efficiency of the path optimization method for a gauge theory}},\ }\href {https://doi.org/10.1103/PhysRevD.107.034509} {\bibfield  {journal} {\bibinfo  {journal} {Phys. Rev. D}\ }\textbf {\bibinfo {volume} {107}},\ \bibinfo {pages} {034509} (\bibinfo {year} {2023})},\ \Eprint {https://arxiv.org/abs/2210.05402} {arXiv:2210.05402 [hep-lat]} \BibitemShut {NoStop}%
\bibitem [{\citenamefont {Kogut}\ and\ \citenamefont {Susskind}(1975)}]{Kogut:1974ag}%
  \BibitemOpen
  \bibfield  {author} {\bibinfo {author} {\bibfnamefont {J.~B.}\ \bibnamefont {Kogut}}\ and\ \bibinfo {author} {\bibfnamefont {L.}~\bibnamefont {Susskind}},\ }\bibfield  {title} {\bibinfo {title} {{Hamiltonian Formulation of Wilson's Lattice Gauge Theories}},\ }\href {https://doi.org/10.1103/PhysRevD.11.395} {\bibfield  {journal} {\bibinfo  {journal} {Phys. Rev. D}\ }\textbf {\bibinfo {volume} {11}},\ \bibinfo {pages} {395} (\bibinfo {year} {1975})}\BibitemShut {NoStop}%
\bibitem [{\citenamefont {Kogut}(1979)}]{Kogut:1979wt}%
  \BibitemOpen
  \bibfield  {author} {\bibinfo {author} {\bibfnamefont {J.~B.}\ \bibnamefont {Kogut}},\ }\bibfield  {title} {\bibinfo {title} {{An Introduction to Lattice Gauge Theory and Spin Systems}},\ }\href {https://doi.org/10.1103/RevModPhys.51.659} {\bibfield  {journal} {\bibinfo  {journal} {Rev. Mod. Phys.}\ }\textbf {\bibinfo {volume} {51}},\ \bibinfo {pages} {659} (\bibinfo {year} {1979})}\BibitemShut {NoStop}%
\bibitem [{\citenamefont {Raychowdhury}\ and\ \citenamefont {Stryker}(2020{\natexlab{a}})}]{Raychowdhury:2019iki}%
  \BibitemOpen
  \bibfield  {author} {\bibinfo {author} {\bibfnamefont {I.}~\bibnamefont {Raychowdhury}}\ and\ \bibinfo {author} {\bibfnamefont {J.~R.}\ \bibnamefont {Stryker}},\ }\bibfield  {title} {\bibinfo {title} {{Loop, string, and hadron dynamics in SU(2) Hamiltonian lattice gauge theories}},\ }\href {https://doi.org/10.1103/PhysRevD.101.114502} {\bibfield  {journal} {\bibinfo  {journal} {Phys. Rev. D}\ }\textbf {\bibinfo {volume} {101}},\ \bibinfo {pages} {114502} (\bibinfo {year} {2020}{\natexlab{a}})},\ \Eprint {https://arxiv.org/abs/1912.06133} {arXiv:1912.06133 [hep-lat]} \BibitemShut {NoStop}%
\bibitem [{\citenamefont {Raychowdhury}\ and\ \citenamefont {Stryker}(2020{\natexlab{b}})}]{Raychowdhury:2018osk}%
  \BibitemOpen
  \bibfield  {author} {\bibinfo {author} {\bibfnamefont {I.}~\bibnamefont {Raychowdhury}}\ and\ \bibinfo {author} {\bibfnamefont {J.~R.}\ \bibnamefont {Stryker}},\ }\bibfield  {title} {\bibinfo {title} {{Solving Gauss's Law on Digital Quantum Computers with Loop-String-Hadron Digitization}},\ }\href {https://doi.org/10.1103/PhysRevResearch.2.033039} {\bibfield  {journal} {\bibinfo  {journal} {Phys. Rev. Res.}\ }\textbf {\bibinfo {volume} {2}},\ \bibinfo {pages} {033039} (\bibinfo {year} {2020}{\natexlab{b}})},\ \Eprint {https://arxiv.org/abs/1812.07554} {arXiv:1812.07554 [hep-lat]} \BibitemShut {NoStop}%
\bibitem [{\citenamefont {Davoudi}\ \emph {et~al.}(2021)\citenamefont {Davoudi}, \citenamefont {Raychowdhury},\ and\ \citenamefont {Shaw}}]{Davoudi:2020yln}%
  \BibitemOpen
  \bibfield  {author} {\bibinfo {author} {\bibfnamefont {Z.}~\bibnamefont {Davoudi}}, \bibinfo {author} {\bibfnamefont {I.}~\bibnamefont {Raychowdhury}},\ and\ \bibinfo {author} {\bibfnamefont {A.}~\bibnamefont {Shaw}},\ }\bibfield  {title} {\bibinfo {title} {{Search for efficient formulations for Hamiltonian simulation of non-Abelian lattice gauge theories}},\ }\href {https://doi.org/10.1103/PhysRevD.104.074505} {\bibfield  {journal} {\bibinfo  {journal} {Phys. Rev. D}\ }\textbf {\bibinfo {volume} {104}},\ \bibinfo {pages} {074505} (\bibinfo {year} {2021})},\ \Eprint {https://arxiv.org/abs/2009.11802} {arXiv:2009.11802 [hep-lat]} \BibitemShut {NoStop}%
\bibitem [{\citenamefont {Zache}\ \emph {et~al.}(2023)\citenamefont {Zache}, \citenamefont {Gonz\'alez-Cuadra},\ and\ \citenamefont {Zoller}}]{Zache:2023dko}%
  \BibitemOpen
  \bibfield  {author} {\bibinfo {author} {\bibfnamefont {T.~V.}\ \bibnamefont {Zache}}, \bibinfo {author} {\bibfnamefont {D.}~\bibnamefont {Gonz\'alez-Cuadra}},\ and\ \bibinfo {author} {\bibfnamefont {P.}~\bibnamefont {Zoller}},\ }\bibfield  {title} {\bibinfo {title} {{Quantum and Classical Spin-Network Algorithms for q-Deformed Kogut-Susskind Gauge Theories}},\ }\href {https://doi.org/10.1103/PhysRevLett.131.171902} {\bibfield  {journal} {\bibinfo  {journal} {Phys. Rev. Lett.}\ }\textbf {\bibinfo {volume} {131}},\ \bibinfo {pages} {171902} (\bibinfo {year} {2023})},\ \Eprint {https://arxiv.org/abs/2304.02527} {arXiv:2304.02527 [quant-ph]} \BibitemShut {NoStop}%
\bibitem [{\citenamefont {Hayata}\ and\ \citenamefont {Hidaka}(2023{\natexlab{a}})}]{Hayata:2023bgh}%
  \BibitemOpen
  \bibfield  {author} {\bibinfo {author} {\bibfnamefont {T.}~\bibnamefont {Hayata}}\ and\ \bibinfo {author} {\bibfnamefont {Y.}~\bibnamefont {Hidaka}},\ }\bibfield  {title} {\bibinfo {title} {{q deformed formulation of Hamiltonian SU(3) Yang-Mills theory}},\ }\href {https://doi.org/10.1007/JHEP09(2023)123} {\bibfield  {journal} {\bibinfo  {journal} {JHEP}\ }\textbf {\bibinfo {volume} {09}},\ \bibinfo {pages} {123}},\ \Eprint {https://arxiv.org/abs/2306.12324} {arXiv:2306.12324 [hep-lat]} \BibitemShut {NoStop}%
\bibitem [{\citenamefont {Hayata}\ and\ \citenamefont {Hidaka}(2023{\natexlab{b}})}]{Hayata:2023puo}%
  \BibitemOpen
  \bibfield  {author} {\bibinfo {author} {\bibfnamefont {T.}~\bibnamefont {Hayata}}\ and\ \bibinfo {author} {\bibfnamefont {Y.}~\bibnamefont {Hidaka}},\ }\bibfield  {title} {\bibinfo {title} {{String-net formulation of Hamiltonian lattice Yang-Mills theories and quantum many-body scars in a nonabelian gauge theory}},\ }\href {https://doi.org/10.1007/JHEP09(2023)126} {\bibfield  {journal} {\bibinfo  {journal} {JHEP}\ }\textbf {\bibinfo {volume} {09}},\ \bibinfo {pages} {126}},\ \Eprint {https://arxiv.org/abs/2305.05950} {arXiv:2305.05950 [hep-lat]} \BibitemShut {NoStop}%
\bibitem [{\citenamefont {Klco}\ \emph {et~al.}(2020)\citenamefont {Klco}, \citenamefont {Stryker},\ and\ \citenamefont {Savage}}]{Klco:2019evd}%
  \BibitemOpen
  \bibfield  {author} {\bibinfo {author} {\bibfnamefont {N.}~\bibnamefont {Klco}}, \bibinfo {author} {\bibfnamefont {J.~R.}\ \bibnamefont {Stryker}},\ and\ \bibinfo {author} {\bibfnamefont {M.~J.}\ \bibnamefont {Savage}},\ }\bibfield  {title} {\bibinfo {title} {{SU(2) non-Abelian gauge field theory in one dimension on digital quantum computers}},\ }\href {https://doi.org/10.1103/PhysRevD.101.074512} {\bibfield  {journal} {\bibinfo  {journal} {Phys. Rev. D}\ }\textbf {\bibinfo {volume} {101}},\ \bibinfo {pages} {074512} (\bibinfo {year} {2020})},\ \Eprint {https://arxiv.org/abs/1908.06935} {arXiv:1908.06935 [quant-ph]} \BibitemShut {NoStop}%
\bibitem [{\citenamefont {Ciavarella}\ \emph {et~al.}(2021)\citenamefont {Ciavarella}, \citenamefont {Klco},\ and\ \citenamefont {Savage}}]{Ciavarella:2021nmj}%
  \BibitemOpen
  \bibfield  {author} {\bibinfo {author} {\bibfnamefont {A.}~\bibnamefont {Ciavarella}}, \bibinfo {author} {\bibfnamefont {N.}~\bibnamefont {Klco}},\ and\ \bibinfo {author} {\bibfnamefont {M.~J.}\ \bibnamefont {Savage}},\ }\bibfield  {title} {\bibinfo {title} {{Trailhead for quantum simulation of SU(3) Yang-Mills lattice gauge theory in the local multiplet basis}},\ }\href {https://doi.org/10.1103/PhysRevD.103.094501} {\bibfield  {journal} {\bibinfo  {journal} {Phys. Rev. D}\ }\textbf {\bibinfo {volume} {103}},\ \bibinfo {pages} {094501} (\bibinfo {year} {2021})},\ \Eprint {https://arxiv.org/abs/2101.10227} {arXiv:2101.10227 [quant-ph]} \BibitemShut {NoStop}%
\bibitem [{\citenamefont {Banerjee}\ \emph {et~al.}(2018)\citenamefont {Banerjee}, \citenamefont {Jiang}, \citenamefont {Olesen}, \citenamefont {Orland},\ and\ \citenamefont {Wiese}}]{Banerjee:2017tjn}%
  \BibitemOpen
  \bibfield  {author} {\bibinfo {author} {\bibfnamefont {D.}~\bibnamefont {Banerjee}}, \bibinfo {author} {\bibfnamefont {F.~J.}\ \bibnamefont {Jiang}}, \bibinfo {author} {\bibfnamefont {T.~Z.}\ \bibnamefont {Olesen}}, \bibinfo {author} {\bibfnamefont {P.}~\bibnamefont {Orland}},\ and\ \bibinfo {author} {\bibfnamefont {U.~J.}\ \bibnamefont {Wiese}},\ }\bibfield  {title} {\bibinfo {title} {{From the $SU(2)$ quantum link model on the honeycomb lattice to the quantum dimer model on the kagom\'e lattice: Phase transition and fractionalized flux strings}},\ }\href {https://doi.org/10.1103/PhysRevB.97.205108} {\bibfield  {journal} {\bibinfo  {journal} {Phys. Rev. B}\ }\textbf {\bibinfo {volume} {97}},\ \bibinfo {pages} {205108} (\bibinfo {year} {2018})},\ \Eprint {https://arxiv.org/abs/1712.08300} {arXiv:1712.08300 [cond-mat.str-el]} \BibitemShut {NoStop}%
\bibitem [{\citenamefont {Byrnes}\ \emph {et~al.}(2002)\citenamefont {Byrnes}, \citenamefont {Sriganesh}, \citenamefont {Bursill},\ and\ \citenamefont {Hamer}}]{Byrnes:2002nv}%
  \BibitemOpen
  \bibfield  {author} {\bibinfo {author} {\bibfnamefont {T.}~\bibnamefont {Byrnes}}, \bibinfo {author} {\bibfnamefont {P.}~\bibnamefont {Sriganesh}}, \bibinfo {author} {\bibfnamefont {R.~J.}\ \bibnamefont {Bursill}},\ and\ \bibinfo {author} {\bibfnamefont {C.~J.}\ \bibnamefont {Hamer}},\ }\bibfield  {title} {\bibinfo {title} {{Density matrix renormalization group approach to the massive Schwinger model}},\ }\href {https://doi.org/10.1103/PhysRevD.66.013002} {\bibfield  {journal} {\bibinfo  {journal} {Phys. Rev. D}\ }\textbf {\bibinfo {volume} {66}},\ \bibinfo {pages} {013002} (\bibinfo {year} {2002})},\ \Eprint {https://arxiv.org/abs/hep-lat/0202014} {arXiv:hep-lat/0202014} \BibitemShut {NoStop}%
\bibitem [{\citenamefont {Ba\~nuls}\ \emph {et~al.}(2013)\citenamefont {Ba\~nuls}, \citenamefont {Cichy}, \citenamefont {Jansen},\ and\ \citenamefont {Cirac}}]{Banuls:2013jaa}%
  \BibitemOpen
  \bibfield  {author} {\bibinfo {author} {\bibfnamefont {M.~C.}\ \bibnamefont {Ba\~nuls}}, \bibinfo {author} {\bibfnamefont {K.}~\bibnamefont {Cichy}}, \bibinfo {author} {\bibfnamefont {K.}~\bibnamefont {Jansen}},\ and\ \bibinfo {author} {\bibfnamefont {J.~I.}\ \bibnamefont {Cirac}},\ }\bibfield  {title} {\bibinfo {title} {{The mass spectrum of the Schwinger model with Matrix Product States}},\ }\href {https://doi.org/10.1007/JHEP11(2013)158} {\bibfield  {journal} {\bibinfo  {journal} {JHEP}\ }\textbf {\bibinfo {volume} {11}},\ \bibinfo {pages} {158}},\ \Eprint {https://arxiv.org/abs/1305.3765} {arXiv:1305.3765 [hep-lat]} \BibitemShut {NoStop}%
\bibitem [{\citenamefont {Martinez}\ \emph {et~al.}(2016)\citenamefont {Martinez} \emph {et~al.}}]{Martinez:2016yna}%
  \BibitemOpen
  \bibfield  {author} {\bibinfo {author} {\bibfnamefont {E.~A.}\ \bibnamefont {Martinez}} \emph {et~al.},\ }\bibfield  {title} {\bibinfo {title} {{Real-time dynamics of lattice gauge theories with a few-qubit quantum computer}},\ }\href {https://doi.org/10.1038/nature18318} {\bibfield  {journal} {\bibinfo  {journal} {Nature}\ }\textbf {\bibinfo {volume} {534}},\ \bibinfo {pages} {516} (\bibinfo {year} {2016})},\ \Eprint {https://arxiv.org/abs/1605.04570} {arXiv:1605.04570 [quant-ph]} \BibitemShut {NoStop}%
\bibitem [{\citenamefont {Muschik}\ \emph {et~al.}(2017)\citenamefont {Muschik}, \citenamefont {Heyl}, \citenamefont {Martinez}, \citenamefont {Monz}, \citenamefont {Schindler}, \citenamefont {Vogell}, \citenamefont {Dalmonte}, \citenamefont {Hauke}, \citenamefont {Blatt},\ and\ \citenamefont {Zoller}}]{Muschik:2016tws}%
  \BibitemOpen
  \bibfield  {author} {\bibinfo {author} {\bibfnamefont {C.}~\bibnamefont {Muschik}}, \bibinfo {author} {\bibfnamefont {M.}~\bibnamefont {Heyl}}, \bibinfo {author} {\bibfnamefont {E.}~\bibnamefont {Martinez}}, \bibinfo {author} {\bibfnamefont {T.}~\bibnamefont {Monz}}, \bibinfo {author} {\bibfnamefont {P.}~\bibnamefont {Schindler}}, \bibinfo {author} {\bibfnamefont {B.}~\bibnamefont {Vogell}}, \bibinfo {author} {\bibfnamefont {M.}~\bibnamefont {Dalmonte}}, \bibinfo {author} {\bibfnamefont {P.}~\bibnamefont {Hauke}}, \bibinfo {author} {\bibfnamefont {R.}~\bibnamefont {Blatt}},\ and\ \bibinfo {author} {\bibfnamefont {P.}~\bibnamefont {Zoller}},\ }\bibfield  {title} {\bibinfo {title} {{U(1) Wilson lattice gauge theories in digital quantum simulators}},\ }\href {https://doi.org/10.1088/1367-2630/aa89ab} {\bibfield  {journal} {\bibinfo  {journal} {New J. Phys.}\ }\textbf {\bibinfo {volume} {19}},\ \bibinfo {pages} {103020} (\bibinfo {year} {2017})},\ \Eprint {https://arxiv.org/abs/1612.08653} {arXiv:1612.08653
  [quant-ph]} \BibitemShut {NoStop}%
\bibitem [{\citenamefont {Klco}\ \emph {et~al.}(2018)\citenamefont {Klco}, \citenamefont {Dumitrescu}, \citenamefont {McCaskey}, \citenamefont {Morris}, \citenamefont {Pooser}, \citenamefont {Sanz}, \citenamefont {Solano}, \citenamefont {Lougovski},\ and\ \citenamefont {Savage}}]{Klco:2018kyo}%
  \BibitemOpen
  \bibfield  {author} {\bibinfo {author} {\bibfnamefont {N.}~\bibnamefont {Klco}}, \bibinfo {author} {\bibfnamefont {E.~F.}\ \bibnamefont {Dumitrescu}}, \bibinfo {author} {\bibfnamefont {A.~J.}\ \bibnamefont {McCaskey}}, \bibinfo {author} {\bibfnamefont {T.~D.}\ \bibnamefont {Morris}}, \bibinfo {author} {\bibfnamefont {R.~C.}\ \bibnamefont {Pooser}}, \bibinfo {author} {\bibfnamefont {M.}~\bibnamefont {Sanz}}, \bibinfo {author} {\bibfnamefont {E.}~\bibnamefont {Solano}}, \bibinfo {author} {\bibfnamefont {P.}~\bibnamefont {Lougovski}},\ and\ \bibinfo {author} {\bibfnamefont {M.~J.}\ \bibnamefont {Savage}},\ }\bibfield  {title} {\bibinfo {title} {{Quantum-classical computation of Schwinger model dynamics using quantum computers}},\ }\href {https://doi.org/10.1103/PhysRevA.98.032331} {\bibfield  {journal} {\bibinfo  {journal} {Phys. Rev. A}\ }\textbf {\bibinfo {volume} {98}},\ \bibinfo {pages} {032331} (\bibinfo {year} {2018})},\ \Eprint {https://arxiv.org/abs/1803.03326} {arXiv:1803.03326 [quant-ph]}
  \BibitemShut {NoStop}%
\bibitem [{\citenamefont {Kokail}\ \emph {et~al.}(2019)\citenamefont {Kokail} \emph {et~al.}}]{Kokail:2018eiw}%
  \BibitemOpen
  \bibfield  {author} {\bibinfo {author} {\bibfnamefont {C.}~\bibnamefont {Kokail}} \emph {et~al.},\ }\bibfield  {title} {\bibinfo {title} {{Self-verifying variational quantum simulation of lattice models}},\ }\href {https://doi.org/10.1038/s41586-019-1177-4} {\bibfield  {journal} {\bibinfo  {journal} {Nature}\ }\textbf {\bibinfo {volume} {569}},\ \bibinfo {pages} {355} (\bibinfo {year} {2019})},\ \Eprint {https://arxiv.org/abs/1810.03421} {arXiv:1810.03421 [quant-ph]} \BibitemShut {NoStop}%
\bibitem [{\citenamefont {Farrell}\ \emph {et~al.}(2024{\natexlab{a}})\citenamefont {Farrell}, \citenamefont {Illa}, \citenamefont {Ciavarella},\ and\ \citenamefont {Savage}}]{Farrell:2023fgd}%
  \BibitemOpen
  \bibfield  {author} {\bibinfo {author} {\bibfnamefont {R.~C.}\ \bibnamefont {Farrell}}, \bibinfo {author} {\bibfnamefont {M.}~\bibnamefont {Illa}}, \bibinfo {author} {\bibfnamefont {A.~N.}\ \bibnamefont {Ciavarella}},\ and\ \bibinfo {author} {\bibfnamefont {M.~J.}\ \bibnamefont {Savage}},\ }\bibfield  {title} {\bibinfo {title} {{Scalable Circuits for Preparing Ground States on Digital Quantum Computers: The Schwinger Model Vacuum on 100 Qubits}},\ }\href {https://doi.org/10.1103/PRXQuantum.5.020315} {\bibfield  {journal} {\bibinfo  {journal} {PRX Quantum}\ }\textbf {\bibinfo {volume} {5}},\ \bibinfo {pages} {020315} (\bibinfo {year} {2024}{\natexlab{a}})},\ \Eprint {https://arxiv.org/abs/2308.04481} {arXiv:2308.04481 [quant-ph]} \BibitemShut {NoStop}%
\bibitem [{\citenamefont {Farrell}\ \emph {et~al.}(2024{\natexlab{b}})\citenamefont {Farrell}, \citenamefont {Illa}, \citenamefont {Ciavarella},\ and\ \citenamefont {Savage}}]{Farrell:2024fit}%
  \BibitemOpen
  \bibfield  {author} {\bibinfo {author} {\bibfnamefont {R.~C.}\ \bibnamefont {Farrell}}, \bibinfo {author} {\bibfnamefont {M.}~\bibnamefont {Illa}}, \bibinfo {author} {\bibfnamefont {A.~N.}\ \bibnamefont {Ciavarella}},\ and\ \bibinfo {author} {\bibfnamefont {M.~J.}\ \bibnamefont {Savage}},\ }\bibfield  {title} {\bibinfo {title} {{Quantum simulations of hadron dynamics in the Schwinger model using 112 qubits}},\ }\href {https://doi.org/10.1103/PhysRevD.109.114510} {\bibfield  {journal} {\bibinfo  {journal} {Phys. Rev. D}\ }\textbf {\bibinfo {volume} {109}},\ \bibinfo {pages} {114510} (\bibinfo {year} {2024}{\natexlab{b}})},\ \Eprint {https://arxiv.org/abs/2401.08044} {arXiv:2401.08044 [quant-ph]} \BibitemShut {NoStop}%
\bibitem [{\citenamefont {Chen}\ \emph {et~al.}(2024)\citenamefont {Chen}, \citenamefont {Yan},\ and\ \citenamefont {Shi}}]{Chen:2024pee}%
  \BibitemOpen
  \bibfield  {author} {\bibinfo {author} {\bibfnamefont {S.}~\bibnamefont {Chen}}, \bibinfo {author} {\bibfnamefont {L.}~\bibnamefont {Yan}},\ and\ \bibinfo {author} {\bibfnamefont {S.}~\bibnamefont {Shi}},\ }\bibfield  {title} {\bibinfo {title} {{Quantum thermalization of Quark-Gluon Plasma}},\ }\href@noop {} {\bibfield  {journal} {\bibinfo  {journal} {\!\!\!}\ } (\bibinfo {year} {2024})},\ \Eprint {https://arxiv.org/abs/2412.00662} {arXiv:2412.00662 [hep-ph]} \BibitemShut {NoStop}%
\bibitem [{\citenamefont {A~Rahman}\ \emph {et~al.}(2021)\citenamefont {A~Rahman}, \citenamefont {Lewis}, \citenamefont {Mendicelli},\ and\ \citenamefont {Powell}}]{ARahman:2021ktn}%
  \BibitemOpen
  \bibfield  {author} {\bibinfo {author} {\bibfnamefont {S.}~\bibnamefont {A~Rahman}}, \bibinfo {author} {\bibfnamefont {R.}~\bibnamefont {Lewis}}, \bibinfo {author} {\bibfnamefont {E.}~\bibnamefont {Mendicelli}},\ and\ \bibinfo {author} {\bibfnamefont {S.}~\bibnamefont {Powell}},\ }\bibfield  {title} {\bibinfo {title} {{SU(2) lattice gauge theory on a quantum annealer}},\ }\href {https://doi.org/10.1103/PhysRevD.104.034501} {\bibfield  {journal} {\bibinfo  {journal} {Phys. Rev. D}\ }\textbf {\bibinfo {volume} {104}},\ \bibinfo {pages} {034501} (\bibinfo {year} {2021})},\ \Eprint {https://arxiv.org/abs/2103.08661} {arXiv:2103.08661 [hep-lat]} \BibitemShut {NoStop}%
\bibitem [{\citenamefont {A~Rahman}\ \emph {et~al.}(2022)\citenamefont {A~Rahman}, \citenamefont {Lewis}, \citenamefont {Mendicelli},\ and\ \citenamefont {Powell}}]{ARahman:2022tkr}%
  \BibitemOpen
  \bibfield  {author} {\bibinfo {author} {\bibfnamefont {S.}~\bibnamefont {A~Rahman}}, \bibinfo {author} {\bibfnamefont {R.}~\bibnamefont {Lewis}}, \bibinfo {author} {\bibfnamefont {E.}~\bibnamefont {Mendicelli}},\ and\ \bibinfo {author} {\bibfnamefont {S.}~\bibnamefont {Powell}},\ }\bibfield  {title} {\bibinfo {title} {{Self-mitigating Trotter circuits for SU(2) lattice gauge theory on a quantum computer}},\ }\href {https://doi.org/10.1103/PhysRevD.106.074502} {\bibfield  {journal} {\bibinfo  {journal} {Phys. Rev. D}\ }\textbf {\bibinfo {volume} {106}},\ \bibinfo {pages} {074502} (\bibinfo {year} {2022})},\ \Eprint {https://arxiv.org/abs/2205.09247} {arXiv:2205.09247 [hep-lat]} \BibitemShut {NoStop}%
\bibitem [{\citenamefont {Hayata}\ and\ \citenamefont {Hidaka}(2021)}]{Hayata:2020xxm}%
  \BibitemOpen
  \bibfield  {author} {\bibinfo {author} {\bibfnamefont {T.}~\bibnamefont {Hayata}}\ and\ \bibinfo {author} {\bibfnamefont {Y.}~\bibnamefont {Hidaka}},\ }\bibfield  {title} {\bibinfo {title} {{Thermalization of Yang-Mills theory in a $(3+1)$ dimensional small lattice system}},\ }\href {https://doi.org/10.1103/PhysRevD.103.094502} {\bibfield  {journal} {\bibinfo  {journal} {Phys. Rev. D}\ }\textbf {\bibinfo {volume} {103}},\ \bibinfo {pages} {094502} (\bibinfo {year} {2021})},\ \Eprint {https://arxiv.org/abs/2011.09814} {arXiv:2011.09814 [hep-lat]} \BibitemShut {NoStop}%
\bibitem [{\citenamefont {Hayata}\ \emph {et~al.}(2021)\citenamefont {Hayata}, \citenamefont {Hidaka},\ and\ \citenamefont {Kikuchi}}]{Hayata:2021kcp}%
  \BibitemOpen
  \bibfield  {author} {\bibinfo {author} {\bibfnamefont {T.}~\bibnamefont {Hayata}}, \bibinfo {author} {\bibfnamefont {Y.}~\bibnamefont {Hidaka}},\ and\ \bibinfo {author} {\bibfnamefont {Y.}~\bibnamefont {Kikuchi}},\ }\bibfield  {title} {\bibinfo {title} {{Diagnosis of information scrambling from Hamiltonian evolution under decoherence}},\ }\href {https://doi.org/10.1103/PhysRevD.104.074518} {\bibfield  {journal} {\bibinfo  {journal} {Phys. Rev. D}\ }\textbf {\bibinfo {volume} {104}},\ \bibinfo {pages} {074518} (\bibinfo {year} {2021})},\ \Eprint {https://arxiv.org/abs/2103.05179} {arXiv:2103.05179 [quant-ph]} \BibitemShut {NoStop}%
\bibitem [{\citenamefont {Yao}(2023)}]{Yao:2023pht}%
  \BibitemOpen
  \bibfield  {author} {\bibinfo {author} {\bibfnamefont {X.}~\bibnamefont {Yao}},\ }\bibfield  {title} {\bibinfo {title} {{SU(2) gauge theory in 2+1 dimensions on a plaquette chain obeys the eigenstate thermalization hypothesis}},\ }\href {https://doi.org/10.1103/PhysRevD.108.L031504} {\bibfield  {journal} {\bibinfo  {journal} {Phys. Rev. D}\ }\textbf {\bibinfo {volume} {108}},\ \bibinfo {pages} {L031504} (\bibinfo {year} {2023})},\ \Eprint {https://arxiv.org/abs/2303.14264} {arXiv:2303.14264 [hep-lat]} \BibitemShut {NoStop}%
\bibitem [{\citenamefont {M\"uller}\ and\ \citenamefont {Yao}(2023)}]{Muller:2023nnk}%
  \BibitemOpen
  \bibfield  {author} {\bibinfo {author} {\bibfnamefont {B.}~\bibnamefont {M\"uller}}\ and\ \bibinfo {author} {\bibfnamefont {X.}~\bibnamefont {Yao}},\ }\bibfield  {title} {\bibinfo {title} {{Simple Hamiltonian for quantum simulation of strongly coupled (2+1)D SU(2) lattice gauge theory on a honeycomb lattice}},\ }\href {https://doi.org/10.1103/PhysRevD.108.094505} {\bibfield  {journal} {\bibinfo  {journal} {Phys. Rev. D}\ }\textbf {\bibinfo {volume} {108}},\ \bibinfo {pages} {094505} (\bibinfo {year} {2023})},\ \Eprint {https://arxiv.org/abs/2307.00045} {arXiv:2307.00045 [quant-ph]} \BibitemShut {NoStop}%
\bibitem [{\citenamefont {Ebner}\ \emph {et~al.}(2024{\natexlab{a}})\citenamefont {Ebner}, \citenamefont {M\"uller}, \citenamefont {Sch\"afer}, \citenamefont {Seidl},\ and\ \citenamefont {Yao}}]{Ebner:2023ixq}%
  \BibitemOpen
  \bibfield  {author} {\bibinfo {author} {\bibfnamefont {L.}~\bibnamefont {Ebner}}, \bibinfo {author} {\bibfnamefont {B.}~\bibnamefont {M\"uller}}, \bibinfo {author} {\bibfnamefont {A.}~\bibnamefont {Sch\"afer}}, \bibinfo {author} {\bibfnamefont {C.}~\bibnamefont {Seidl}},\ and\ \bibinfo {author} {\bibfnamefont {X.}~\bibnamefont {Yao}},\ }\bibfield  {title} {\bibinfo {title} {{Eigenstate thermalization in (2+1)-dimensional SU(2) lattice gauge theory}},\ }\href {https://doi.org/10.1103/PhysRevD.109.014504} {\bibfield  {journal} {\bibinfo  {journal} {Phys. Rev. D}\ }\textbf {\bibinfo {volume} {109}},\ \bibinfo {pages} {014504} (\bibinfo {year} {2024}{\natexlab{a}})},\ \Eprint {https://arxiv.org/abs/2308.16202} {arXiv:2308.16202 [hep-lat]} \BibitemShut {NoStop}%
\bibitem [{\citenamefont {Turro}\ \emph {et~al.}(2024)\citenamefont {Turro}, \citenamefont {Ciavarella},\ and\ \citenamefont {Yao}}]{Turro:2024pxu}%
  \BibitemOpen
  \bibfield  {author} {\bibinfo {author} {\bibfnamefont {F.}~\bibnamefont {Turro}}, \bibinfo {author} {\bibfnamefont {A.}~\bibnamefont {Ciavarella}},\ and\ \bibinfo {author} {\bibfnamefont {X.}~\bibnamefont {Yao}},\ }\bibfield  {title} {\bibinfo {title} {{Classical and quantum computing of shear viscosity for (2+1)D SU(2) gauge theory}},\ }\href {https://doi.org/10.1103/PhysRevD.109.114511} {\bibfield  {journal} {\bibinfo  {journal} {Phys. Rev. D}\ }\textbf {\bibinfo {volume} {109}},\ \bibinfo {pages} {114511} (\bibinfo {year} {2024})},\ \Eprint {https://arxiv.org/abs/2402.04221} {arXiv:2402.04221 [hep-lat]} \BibitemShut {NoStop}%
\bibitem [{\citenamefont {Ebner}\ \emph {et~al.}(2024{\natexlab{b}})\citenamefont {Ebner}, \citenamefont {Sch\"afer}, \citenamefont {Seidl}, \citenamefont {M\"uller},\ and\ \citenamefont {Yao}}]{Ebner:2024mee}%
  \BibitemOpen
  \bibfield  {author} {\bibinfo {author} {\bibfnamefont {L.}~\bibnamefont {Ebner}}, \bibinfo {author} {\bibfnamefont {A.}~\bibnamefont {Sch\"afer}}, \bibinfo {author} {\bibfnamefont {C.}~\bibnamefont {Seidl}}, \bibinfo {author} {\bibfnamefont {B.}~\bibnamefont {M\"uller}},\ and\ \bibinfo {author} {\bibfnamefont {X.}~\bibnamefont {Yao}},\ }\bibfield  {title} {\bibinfo {title} {{Entanglement entropy of (2+1)-dimensional SU(2) lattice gauge theory on plaquette chains}},\ }\href {https://doi.org/10.1103/PhysRevD.110.014505} {\bibfield  {journal} {\bibinfo  {journal} {Phys. Rev. D}\ }\textbf {\bibinfo {volume} {110}},\ \bibinfo {pages} {014505} (\bibinfo {year} {2024}{\natexlab{b}})},\ \Eprint {https://arxiv.org/abs/2401.15184} {arXiv:2401.15184 [hep-lat]} \BibitemShut {NoStop}%
\bibitem [{\citenamefont {Turro}\ and\ \citenamefont {Yao}(2025)}]{Turro:2025sec}%
  \BibitemOpen
  \bibfield  {author} {\bibinfo {author} {\bibfnamefont {F.}~\bibnamefont {Turro}}\ and\ \bibinfo {author} {\bibfnamefont {X.}~\bibnamefont {Yao}},\ }\bibfield  {title} {\bibinfo {title} {{Emergent Hydrodynamic Mode on SU(2) Plaquette Chains and Quantum Simulation}},\ }\href@noop {} {\bibfield  {journal} {\bibinfo  {journal} {\!\!\!}\ } (\bibinfo {year} {2025})},\ \Eprint {https://arxiv.org/abs/2502.17551} {arXiv:2502.17551 [hep-ph]} \BibitemShut {NoStop}%
\bibitem [{\citenamefont {Illa}\ \emph {et~al.}(2025)\citenamefont {Illa}, \citenamefont {Savage},\ and\ \citenamefont {Yao}}]{Illa:2025dou}%
  \BibitemOpen
  \bibfield  {author} {\bibinfo {author} {\bibfnamefont {M.}~\bibnamefont {Illa}}, \bibinfo {author} {\bibfnamefont {M.~J.}\ \bibnamefont {Savage}},\ and\ \bibinfo {author} {\bibfnamefont {X.}~\bibnamefont {Yao}},\ }\bibfield  {title} {\bibinfo {title} {{Improved Honeycomb and Hyper-Honeycomb Lattice Hamiltonians for Quantum Simulations of Non-Abelian Gauge Theories}},\ }\href@noop {} {\bibfield  {journal} {\bibinfo  {journal} {\!\!\!}\ } (\bibinfo {year} {2025})},\ \Eprint {https://arxiv.org/abs/2503.09688} {arXiv:2503.09688 [hep-lat]} \BibitemShut {NoStop}%
\bibitem [{\citenamefont {Heinz}\ and\ \citenamefont {Kolb}(2002)}]{Heinz:2001xi}%
  \BibitemOpen
  \bibfield  {author} {\bibinfo {author} {\bibfnamefont {U.~W.}\ \bibnamefont {Heinz}}\ and\ \bibinfo {author} {\bibfnamefont {P.~F.}\ \bibnamefont {Kolb}},\ }\bibfield  {title} {\bibinfo {title} {{Early thermalization at RHIC}},\ }\href {https://doi.org/10.1016/S0375-9474(02)00714-5} {\bibfield  {journal} {\bibinfo  {journal} {Nucl. Phys. A}\ }\textbf {\bibinfo {volume} {702}},\ \bibinfo {pages} {269} (\bibinfo {year} {2002})},\ \Eprint {https://arxiv.org/abs/hep-ph/0111075} {arXiv:hep-ph/0111075} \BibitemShut {NoStop}%
\bibitem [{\citenamefont {Berges}\ \emph {et~al.}(2021)\citenamefont {Berges}, \citenamefont {Heller}, \citenamefont {Mazeliauskas},\ and\ \citenamefont {Venugopalan}}]{Berges:2020fwq}%
  \BibitemOpen
  \bibfield  {author} {\bibinfo {author} {\bibfnamefont {J.}~\bibnamefont {Berges}}, \bibinfo {author} {\bibfnamefont {M.~P.}\ \bibnamefont {Heller}}, \bibinfo {author} {\bibfnamefont {A.}~\bibnamefont {Mazeliauskas}},\ and\ \bibinfo {author} {\bibfnamefont {R.}~\bibnamefont {Venugopalan}},\ }\bibfield  {title} {\bibinfo {title} {{QCD thermalization: Ab initio approaches and interdisciplinary connections}},\ }\href {https://doi.org/10.1103/RevModPhys.93.035003} {\bibfield  {journal} {\bibinfo  {journal} {Rev. Mod. Phys.}\ }\textbf {\bibinfo {volume} {93}},\ \bibinfo {pages} {035003} (\bibinfo {year} {2021})},\ \Eprint {https://arxiv.org/abs/2005.12299} {arXiv:2005.12299 [hep-th]} \BibitemShut {NoStop}%
\bibitem [{\citenamefont {McLerran}\ and\ \citenamefont {Venugopalan}(1994{\natexlab{a}})}]{McLerran:1993ni}%
  \BibitemOpen
  \bibfield  {author} {\bibinfo {author} {\bibfnamefont {L.~D.}\ \bibnamefont {McLerran}}\ and\ \bibinfo {author} {\bibfnamefont {R.}~\bibnamefont {Venugopalan}},\ }\bibfield  {title} {\bibinfo {title} {{Computing quark and gluon distribution functions for very large nuclei}},\ }\href {https://doi.org/10.1103/PhysRevD.49.2233} {\bibfield  {journal} {\bibinfo  {journal} {Phys. Rev. D}\ }\textbf {\bibinfo {volume} {49}},\ \bibinfo {pages} {2233} (\bibinfo {year} {1994}{\natexlab{a}})},\ \Eprint {https://arxiv.org/abs/hep-ph/9309289} {arXiv:hep-ph/9309289} \BibitemShut {NoStop}%
\bibitem [{\citenamefont {McLerran}\ and\ \citenamefont {Venugopalan}(1994{\natexlab{b}})}]{McLerran:1993ka}%
  \BibitemOpen
  \bibfield  {author} {\bibinfo {author} {\bibfnamefont {L.~D.}\ \bibnamefont {McLerran}}\ and\ \bibinfo {author} {\bibfnamefont {R.}~\bibnamefont {Venugopalan}},\ }\bibfield  {title} {\bibinfo {title} {{Gluon distribution functions for very large nuclei at small transverse momentum}},\ }\href {https://doi.org/10.1103/PhysRevD.49.3352} {\bibfield  {journal} {\bibinfo  {journal} {Phys. Rev. D}\ }\textbf {\bibinfo {volume} {49}},\ \bibinfo {pages} {3352} (\bibinfo {year} {1994}{\natexlab{b}})},\ \Eprint {https://arxiv.org/abs/hep-ph/9311205} {arXiv:hep-ph/9311205} \BibitemShut {NoStop}%
\bibitem [{\citenamefont {McLerran}\ and\ \citenamefont {Venugopalan}(1994{\natexlab{c}})}]{McLerran:1994vd}%
  \BibitemOpen
  \bibfield  {author} {\bibinfo {author} {\bibfnamefont {L.~D.}\ \bibnamefont {McLerran}}\ and\ \bibinfo {author} {\bibfnamefont {R.}~\bibnamefont {Venugopalan}},\ }\bibfield  {title} {\bibinfo {title} {{Green's functions in the color field of a large nucleus}},\ }\href {https://doi.org/10.1103/PhysRevD.50.2225} {\bibfield  {journal} {\bibinfo  {journal} {Phys. Rev. D}\ }\textbf {\bibinfo {volume} {50}},\ \bibinfo {pages} {2225} (\bibinfo {year} {1994}{\natexlab{c}})},\ \Eprint {https://arxiv.org/abs/hep-ph/9402335} {arXiv:hep-ph/9402335} \BibitemShut {NoStop}%
\bibitem [{\citenamefont {Kovner}\ \emph {et~al.}(1995)\citenamefont {Kovner}, \citenamefont {McLerran},\ and\ \citenamefont {Weigert}}]{Kovner:1995ja}%
  \BibitemOpen
  \bibfield  {author} {\bibinfo {author} {\bibfnamefont {A.}~\bibnamefont {Kovner}}, \bibinfo {author} {\bibfnamefont {L.~D.}\ \bibnamefont {McLerran}},\ and\ \bibinfo {author} {\bibfnamefont {H.}~\bibnamefont {Weigert}},\ }\bibfield  {title} {\bibinfo {title} {{Gluon production from nonAbelian Weizsacker-Williams fields in nucleus-nucleus collisions}},\ }\href {https://doi.org/10.1103/PhysRevD.52.6231} {\bibfield  {journal} {\bibinfo  {journal} {Phys. Rev. D}\ }\textbf {\bibinfo {volume} {52}},\ \bibinfo {pages} {6231} (\bibinfo {year} {1995})},\ \Eprint {https://arxiv.org/abs/hep-ph/9502289} {arXiv:hep-ph/9502289} \BibitemShut {NoStop}%
\bibitem [{\citenamefont {Deutsch}(1991)}]{PhysRevA.43.2046}%
  \BibitemOpen
  \bibfield  {author} {\bibinfo {author} {\bibfnamefont {J.~M.}\ \bibnamefont {Deutsch}},\ }\bibfield  {title} {\bibinfo {title} {Quantum statistical mechanics in a closed system},\ }\href {https://doi.org/10.1103/PhysRevA.43.2046} {\bibfield  {journal} {\bibinfo  {journal} {Phys. Rev. A}\ }\textbf {\bibinfo {volume} {43}},\ \bibinfo {pages} {2046} (\bibinfo {year} {1991})}\BibitemShut {NoStop}%
\bibitem [{\citenamefont {Srednicki}(1994)}]{PhysRevE.50.888}%
  \BibitemOpen
  \bibfield  {author} {\bibinfo {author} {\bibfnamefont {M.}~\bibnamefont {Srednicki}},\ }\bibfield  {title} {\bibinfo {title} {Chaos and quantum thermalization},\ }\href {https://doi.org/10.1103/PhysRevE.50.888} {\bibfield  {journal} {\bibinfo  {journal} {Phys. Rev. E}\ }\textbf {\bibinfo {volume} {50}},\ \bibinfo {pages} {888} (\bibinfo {year} {1994})}\BibitemShut {NoStop}%
\bibitem [{\citenamefont {Rigol}\ \emph {et~al.}(2008)\citenamefont {Rigol}, \citenamefont {Dunjko},\ and\ \citenamefont {Olshanii}}]{Rigol:2007juv}%
  \BibitemOpen
  \bibfield  {author} {\bibinfo {author} {\bibfnamefont {M.}~\bibnamefont {Rigol}}, \bibinfo {author} {\bibfnamefont {V.}~\bibnamefont {Dunjko}},\ and\ \bibinfo {author} {\bibfnamefont {M.}~\bibnamefont {Olshanii}},\ }\bibfield  {title} {\bibinfo {title} {{Thermalization and its mechanism for generic isolated quantum systems}},\ }\href {https://doi.org/10.1038/nature06838} {\bibfield  {journal} {\bibinfo  {journal} {Nature}\ }\textbf {\bibinfo {volume} {452}},\ \bibinfo {pages} {854} (\bibinfo {year} {2008})},\ \Eprint {https://arxiv.org/abs/0708.1324} {arXiv:0708.1324 [cond-mat.stat-mech]} \BibitemShut {NoStop}%
\bibitem [{\citenamefont {Cataldi}\ \emph {et~al.}(2024)\citenamefont {Cataldi}, \citenamefont {Magnifico}, \citenamefont {Silvi},\ and\ \citenamefont {Montangero}}]{cataldi2024simulating}%
  \BibitemOpen
  \bibfield  {author} {\bibinfo {author} {\bibfnamefont {G.}~\bibnamefont {Cataldi}}, \bibinfo {author} {\bibfnamefont {G.}~\bibnamefont {Magnifico}}, \bibinfo {author} {\bibfnamefont {P.}~\bibnamefont {Silvi}},\ and\ \bibinfo {author} {\bibfnamefont {S.}~\bibnamefont {Montangero}},\ }\bibfield  {title} {\bibinfo {title} {Simulating (2+ 1) d su (2) yang-mills lattice gauge theory at finite density with tensor networks},\ }\href@noop {} {\bibfield  {journal} {\bibinfo  {journal} {Physical Review Research}\ }\textbf {\bibinfo {volume} {6}},\ \bibinfo {pages} {033057} (\bibinfo {year} {2024})}\BibitemShut {NoStop}%
\bibitem [{\citenamefont {Susskind}(1977)}]{Susskind:1976jm}%
  \BibitemOpen
  \bibfield  {author} {\bibinfo {author} {\bibfnamefont {L.}~\bibnamefont {Susskind}},\ }\bibfield  {title} {\bibinfo {title} {{Lattice Fermions}},\ }\href {https://doi.org/10.1103/PhysRevD.16.3031} {\bibfield  {journal} {\bibinfo  {journal} {Phys. Rev. D}\ }\textbf {\bibinfo {volume} {16}},\ \bibinfo {pages} {3031} (\bibinfo {year} {1977})}\BibitemShut {NoStop}%
\bibitem [{\citenamefont {Nielsen}\ and\ \citenamefont {Ninomiya}(1981{\natexlab{a}})}]{Nielsen:1980rz}%
  \BibitemOpen
  \bibfield  {author} {\bibinfo {author} {\bibfnamefont {H.~B.}\ \bibnamefont {Nielsen}}\ and\ \bibinfo {author} {\bibfnamefont {M.}~\bibnamefont {Ninomiya}},\ }\bibfield  {title} {\bibinfo {title} {{Absence of Neutrinos on a Lattice. 1. Proof by Homotopy Theory}},\ }\href {https://doi.org/10.1016/0550-3213(82)90011-6} {\bibfield  {journal} {\bibinfo  {journal} {Nucl. Phys. B}\ }\textbf {\bibinfo {volume} {185}},\ \bibinfo {pages} {20} (\bibinfo {year} {1981}{\natexlab{a}})},\ \bibinfo {note} {[Erratum: Nucl.Phys.B 195, 541 (1982)]}\BibitemShut {NoStop}%
\bibitem [{\citenamefont {Nielsen}\ and\ \citenamefont {Ninomiya}(1981{\natexlab{b}})}]{Nielsen:1981xu}%
  \BibitemOpen
  \bibfield  {author} {\bibinfo {author} {\bibfnamefont {H.~B.}\ \bibnamefont {Nielsen}}\ and\ \bibinfo {author} {\bibfnamefont {M.}~\bibnamefont {Ninomiya}},\ }\bibfield  {title} {\bibinfo {title} {{Absence of Neutrinos on a Lattice. 2. Intuitive Topological Proof}},\ }\href {https://doi.org/10.1016/0550-3213(81)90524-1} {\bibfield  {journal} {\bibinfo  {journal} {Nucl. Phys. B}\ }\textbf {\bibinfo {volume} {193}},\ \bibinfo {pages} {173} (\bibinfo {year} {1981}{\natexlab{b}})}\BibitemShut {NoStop}%
\bibitem [{\citenamefont {Mathur}(2005)}]{Mathur:2004kr}%
  \BibitemOpen
  \bibfield  {author} {\bibinfo {author} {\bibfnamefont {M.}~\bibnamefont {Mathur}},\ }\bibfield  {title} {\bibinfo {title} {{Harmonic oscillator prepotentials in SU(2) lattice gauge theory}},\ }\href {https://doi.org/10.1088/0305-4470/38/46/008} {\bibfield  {journal} {\bibinfo  {journal} {J. Phys. A}\ }\textbf {\bibinfo {volume} {38}},\ \bibinfo {pages} {10015} (\bibinfo {year} {2005})},\ \Eprint {https://arxiv.org/abs/hep-lat/0403029} {arXiv:hep-lat/0403029} \BibitemShut {NoStop}%
\bibitem [{\citenamefont {P}\ and\ \citenamefont {Anishetty}(2021{\natexlab{a}})}]{P:2019qdq}%
  \BibitemOpen
  \bibfield  {author} {\bibinfo {author} {\bibfnamefont {S.~T.}\ \bibnamefont {P}}\ and\ \bibinfo {author} {\bibfnamefont {R.}~\bibnamefont {Anishetty}},\ }\bibfield  {title} {\bibinfo {title} {{Gauss law in lattice QCD and its gauge-invariant Hilbert space}},\ }\href {https://doi.org/10.1007/s12648-021-02142-w} {\bibfield  {journal} {\bibinfo  {journal} {Indian J. Phys.}\ }\textbf {\bibinfo {volume} {95}},\ \bibinfo {pages} {1651} (\bibinfo {year} {2021}{\natexlab{a}})},\ \Eprint {https://arxiv.org/abs/1906.03893} {arXiv:1906.03893 [hep-lat]} \BibitemShut {NoStop}%
\bibitem [{\citenamefont {Trotter}(1959)}]{trotter1959product}%
  \BibitemOpen
  \bibfield  {author} {\bibinfo {author} {\bibfnamefont {H.~F.}\ \bibnamefont {Trotter}},\ }\bibfield  {title} {\bibinfo {title} {On the product of semi-groups of operators},\ }\href@noop {} {\bibfield  {journal} {\bibinfo  {journal} {Proceedings of the American Mathematical Society}\ }\textbf {\bibinfo {volume} {10}},\ \bibinfo {pages} {545} (\bibinfo {year} {1959})}\BibitemShut {NoStop}%
\bibitem [{\citenamefont {Suzuki}(1976)}]{suzuki1976generalized}%
  \BibitemOpen
  \bibfield  {author} {\bibinfo {author} {\bibfnamefont {M.}~\bibnamefont {Suzuki}},\ }\bibfield  {title} {\bibinfo {title} {Generalized trotter's formula and systematic approximants of exponential operators and inner derivations with applications to many-body problems},\ }\href@noop {} {\bibfield  {journal} {\bibinfo  {journal} {Communications in Mathematical Physics}\ }\textbf {\bibinfo {volume} {51}},\ \bibinfo {pages} {183} (\bibinfo {year} {1976})}\BibitemShut {NoStop}%
\bibitem [{\citenamefont {Javadi-Abhari}\ \emph {et~al.}(2024)\citenamefont {Javadi-Abhari} \emph {et~al.}}]{Javadi-Abhari:2024kbf}%
  \BibitemOpen
  \bibfield  {author} {\bibinfo {author} {\bibfnamefont {A.}~\bibnamefont {Javadi-Abhari}} \emph {et~al.},\ }\href@noop {} {\bibinfo {title} {{Quantum computing with Qiskit}}} (\bibinfo {year} {2024}),\ \Eprint {https://arxiv.org/abs/2405.08810} {arXiv:2405.08810 [quant-ph]} \BibitemShut {NoStop}%
\bibitem [{\citenamefont {P}\ and\ \citenamefont {Anishetty}(2021{\natexlab{b}})}]{anishetty2019gauss}%
  \BibitemOpen
  \bibfield  {author} {\bibinfo {author} {\bibfnamefont {S.~T.}\ \bibnamefont {P}}\ and\ \bibinfo {author} {\bibfnamefont {R.}~\bibnamefont {Anishetty}},\ }\bibfield  {title} {\bibinfo {title} {{Gauss law in lattice QCD and its gauge-invariant Hilbert space}},\ }\href {https://doi.org/10.1007/s12648-021-02142-w} {\bibfield  {journal} {\bibinfo  {journal} {Indian J. Phys.}\ }\textbf {\bibinfo {volume} {95}},\ \bibinfo {pages} {1651} (\bibinfo {year} {2021}{\natexlab{b}})},\ \Eprint {https://arxiv.org/abs/1906.03893} {arXiv:1906.03893 [hep-lat]} \BibitemShut {NoStop}%
\end{thebibliography}%
				
			\end{document}